\documentclass[]{pasj02} 
\usepackage[switch,mathlines]{lineno} % add line number to manuscript

\jyear{2025}
\Received{}%{yyyy/mm/dd}
\Accepted{}%{yyyy/mm/dd}
\usepackage{xcolor}
\usepackage{diagbox}

\begin{document} 

\title{An optical transient candidate of $\lesssim$ 2-second duration captured by wide-field video observations}

%%% begin:list of authors
% Do NOT capitalize all letters in "textsc".
\author{
Noriaki \textsc{Arima},\altaffilmark{1,2}$^{,*}$\orcid{0000-0002-2721-7109} \email{arima.noriaki@nihon-u.ac.jp}
Mamoru \textsc{Doi},\altaffilmark{2,3}
Shigeyuki \textsc{Sako},\altaffilmark{2,4,5} \orcid{0000-0002-8792-2205}
Yuu \textsc{Niino},\altaffilmark{6} \orcid{0000-0001-5322-5076}
Ryou \textsc{Ohsawa},\altaffilmark{3,7} \orcid{0000-0001-5797-6010}
Nozomu \textsc{Tominaga},\altaffilmark{3,7,8}
Masaomi \textsc{Tanaka},\altaffilmark{9} \orcid{0000-0001-8253-6850}
Michael \textsc{Richmond},\altaffilmark{10} \orcid{0000-0001-8537-3153}
Shinsuke \textsc{Abe},\altaffilmark{1} \orcid{0000-0002-6247-4979}
Naoto \textsc{Kobayashi},\altaffilmark{6} \orcid{0000-0003-4578-2619}
Sohei \textsc{Kondo},\altaffilmark{6} 
Yuki \textsc{Mori},\altaffilmark{6}
Ko \textsc{Arimatsu},\altaffilmark{11,12} \orcid{0000-0003-1260-9502}
Toshihiro \textsc{Kasuga},\altaffilmark{3,13} \orcid{0000-0001-5903-7391}
Shin-ichiro \textsc{Okumura},\altaffilmark{14} \orcid{0000-0002-1873-3494}
Jun-ichi \textsc{Watanabe},\altaffilmark{3,7} \orcid{0000-0003-4391-4446}
and
Takuya \textsc{Yamashita}\altaffilmark{3}
}
%%%
\altaffiltext{1}{Department of Aerospace Engineering, Nihon University, 7-24-1 Narashinodai, Funabashi, Chiba, 274-8501, Japan}
\altaffiltext{2}{Institute of Astronomy, Graduate School of Science, The University of Tokyo, 2-21-1 Osawa, Mitaka, Tokyo 181-0015, Japan}
\altaffiltext{3}{National Astronomical Observatory of Japan, National Institutes of Natural Sciences, 2-21-1 Osawa, Mitaka, Tokyo 181-8588, Japan}
\altaffiltext{4}{UTokyo Organization for Planetary Space Science, The University of Tokyo, 7-3-1 Hongo, Bunkyo-ku, Tokyo 113-0033, Japan}
\altaffiltext{5}{Collaborative Research Organization for Space Science and Technology, The University of Tokyo, 7-3-1 Hongo, Bunkyo-ku, Tokyo 113-0033, Japan}
\altaffiltext{6}{Kiso Observatory, Institute of Astronomy, Graduate School of Science, The University of Tokyo, 10762-30 Mitake, Kiso-machi, Kiso-gun, Nagano 397-0101, Japan}
\altaffiltext{7}{Astronomical Science Program, Graduate Institute for Advanced Studies, SOKENDAI 2-21-1 Osawa, Mitaka, Tokyo 181-8588, Japan}
\altaffiltext{8}{Department of Physics, Faculty of Science and Engineering, Konan University, 8-9-1 Okamoto, Kobe, Hyogo 658-8501, Japan}
\altaffiltext{9}{Astronomical Institute, Tohoku University, Sendai 980-8578, Japan}
\altaffiltext{10}{Physics Department, Rochester Institute of Technology, 84 Lomb Memorial Drive, Rochester, NY 14623, USA}
\altaffiltext{11}{Astronomical Observatory, Graduate School of Science, Kyoto University, Kitashirakawa-oiwake-cho, Sakyo-ku, Kyoto, Kyoto 606-8502, Japan}
\altaffiltext{12}{Ishigakijima Astronomical Observatory, 1024-1, Arakawa, Ishigaki, Okinawa, 907-0024, Japan}
\altaffiltext{13}{Nishi-Harima Astronomical Observatory, Center for Astronomy, University of Hyogo, 407-2, Nishigaichi, Sayo, Hyogo, 679-5313, Japan}
\altaffiltext{14}{Japan Spaceguard Association, Bisei Spaceguard Center, 1716-3 Okura, Bisei, Ibara, Okayama 714-1411, Japan}

\footnotetext[$*$]{Corresponding author. Email: arima.noriaki@nihon-u.ac.jp}

%%% end:list of authors

%% !!! Select 3 to 5 words from PASJ's key words !!! 
%% List of Key Words: https://academic.oup.com/pasj/pages/Pasj_Keywords 
%% "\KeyWords{ }" always has to be placed before ``\maketitle'' 
% \KeyWords{xxxx: xxxx --- ......} 

\maketitle

\maketitle

% % For arXiv submission purpose
%----------------------------------------------
% --- 1ページ目フッタの高さ不足を修正（pasj02.cls対策） ---
% \makeatletter
% \setlength{\footskip}{28pt} % ← まずは 24–32pt の範囲で調整
% \def\@oddfoot{%
%   \hbox to \textwidth{%
%     \vbox{%
%       \hsize\textwidth
%       \rule{\textwidth}{.5pt}\par\vspace*{1pt}%
%       {\historyfont
%         \textbf{Received:}\ \@make@formatted@date\rdate,
%         \textbf{Accepted:}\ \@make@formatted@date\adate\par}%
%       {\rffont\copyright\ \@jyear.\ Astronomical Society of Japan.\par}%
%     }%
%   }%
% }%
% \makeatother

\begin{abstract}
Recent time-domain surveys in the optical have revealed rapid transients that evolve on timescales of $\lesssim 10$ days, expanding the population of transients toward the short-duration regime. The transient search on even shorter timescales, particularly those lasting only seconds, or less, remains a largely unexplored frontier, offering the significant potential for discovering objects from unexpected populations. Very short-duration optical transients could be potential counterparts to millisecond-duration fast radio bursts (FRBs), 
providing clues about their origins. However, the optical search for transients on such short timescales has been limited primarily due to instrumental constraints. 
Here we report the discovery of an optical transient candidate (TMG20200322) with a duration of $\lesssim 2\ {\rm s}$ by wide-field video observations in the direction of the Earth's shadow. TMG20200322 was detected in just two consecutive images of 1-second exposure time, with its shape becoming elongated in the second frame. PSF shape variability analysis of the field stars reveals such an elongated PSF cannot be explained by atmospheric fluctuations. 
We investigate the potential origins of TMG20200322 in two scenarios: meteoroid impact flashes on near-Earth asteroids (NEAs) and head-on meteors in the Earth’s atmosphere. None of the scenarios provides a satisfactory explanation for this transient. We derive a sky-projected rate of the TMG20200322 event to be $R_{\rm trans} = (3.4 \times 10^{-2})^{+0.13}_{-0.028} \ {\rm deg^{-2}\ day^{-1}}$ and an upper limit of second-timescale transients with durations of $1 \rm \ s \leq \tau \lesssim 15 \rm \ s$ to be $R_{\rm trans} \lesssim 0.10\ {\rm deg^{-2}\ day^{-1}}$ for the non-detection case. 
We highlight that continuous monitoring observations in the direction of the Earth's shadow could be a key strategy to unveiling a new population of optical transients on timescales of seconds or less. 
\end{abstract}

\pagewiselinenumbers 
\nolinenumbers

\section{Introduction}\label{sec:intro}
In recent years, time-domain sky surveys have dramatically expanded our ability to detect and classify transients of various kinds. By visiting the same field once every few days or less, high-cadence optical surveys such as All-Sky Automated Survey for   
\newpage \noindent
Supernovae (ASAS-SN; \cite{Shappee2014}),
the Asteroid Terrestrial-impact Last Alert System (ATLAS; \cite{Tonry2018,Smith2020}), the The Panoramic Survey Telescope and Rapid Response System (Pan-STARRS; \cite{Chambers2016}) and the Zwicky Transient Facility (ZTF; \cite{Bellm2019}) have led to the identification of a diverse set of rapidly evolving transients. 
These rapid transients can reach peak luminosities comparable or even exceeding those of typical supernovae, yet their thermal emission that evolves on much shorter ($\lesssim 10$ days) timescales (e.g., \cite{Drout2014,Arcavi2016,Prentice2018,Ho2020}). The Vera C. Rubin Observatory's Legacy Survey of Space and Time 
(LSST; \cite{Ivezic2019}), which will soon begin science operations, will reveal the 
dynamic sky by discovering millions of transient phenomena nightly. 
\par
The discovery of new populations of transients often occurs serendipitously when exploring new regions of observational parameter space. Gamma-ray bursts (GRBs) are now recognized as $\gamma$-ray transients with durations ranging from sub-seconds to hundreds of seconds and powered by relativistic jets of extragalactic origin \citep{Meegan1992}. GRBs were first discovered in the late 1960s by the military Vela satellites, which were originally designed to monitor $\gamma$-rays associated with nuclear explosions in outer space \citep{Klebesadel1973}. A handful of observations have successfully detected the prompt optical emission for some GRBs (e.g., \cite{Vestrand2005,Racusin2008}). 
A millisecond-duration radio pulse—now known as a fast radio burst (FRB)—was first identified in archival 1.4 GHz radio pulsar survey data and reported in 2007 \citep{Lorimer2007}. Although FRBs are now widely recognized as a new class of extragalactic radio transients, their origin and physical mechanisms responsible for the brief, high-energy radio emission are not well understood \citep{Thornton2013,Petroff2016,Platts2019}. So far, no extragalactic optical transient counterparts to FRBs have been identified, with the only exception of the X-ray burst from the Galactic magnetar SGR J1935+2154, which was associated with FRB 200428 \citep{Bochenek2020,Li2021}.
\par
Continuous imaging of the same sky region allows us to search for transient phenomena on the shortest timescale, set by the exposure time of individual frames. \citet{Berger2013} searched for fast optical transients with durations as short as approximately 33 minutes by the Pan-STARRS Medium-Deep Survey using consecutive two photometric $g,\ r$-bands. Although several flare events of Galactic M-dwarf stars were detected, no plausible fast extragalactic transients were found. With the non-detection result, they placed an upper limit on the sky-projected event rate of extragalactic fast transients at $\lesssim 22.5$ mag of $R(\tau = 33\ {\rm min}) \lesssim 0.12\ \deg^{-2}\ {\rm day^{-1}}$. With detectable timescales of 1.2 minutes and limiting magnitude of $m_g \sim 23$ mag, \citet{Andreoni2020} conducted a multi-wavelength (from $\gamma-$rays to radio bands) transient search for the primary purpose of discovering optical counterparts to FRBs. Nine candidates were discovered in the program, all of which were classified as probably stellar flares on the basis of their multiband images and follow-up observations, giving an upper limit of the event rate of transients of $R(\tau = 1.2\ {\rm min}) \lesssim 1.6\ \deg^{-2}\ {\rm day^{-1}}$. 
\par 
The parameter space of optical transients lasting a few seconds or 
less remains largely unexplored. To date, only a handful of optical surveys have systematically attempted to probe timescales of a few seconds or shorter, and very few have simultaneously achieved both moderate sensitivity and a wide field of view (FoV). The typical readout time of CCDs, on the order of $\sim$10 seconds, constitutes one of the major limitations in probing the second-timescale regime. As a result, our understanding of the occurrence rates and spatial distributions of optical phenomena on timescales of a few seconds or less remains limited. In particular, models such as the magnetar flare scenario for FRBs predict optical flashes lasting less than one second (\cite{Yang2019,Chen2020,Beloborodov2020}), yet remain observationally elusive. The “Pi of the Sky” project \citep{Sokolowski2010} performed a transient search with a timescale of $\gtrsim 10$ seconds and a depth of $11$ mag, giving an all-sky upper limit of transients brighter than 11 mag is $< 300$ per day. For ultra-fast ($\ll 1$ s) time domain searches, \citet{Tingay2021} performed observations with dual small camera systems (diameter of $ \lesssim 10$ cm) and an exposure time of 21 ms, reaching a limiting magnitude of $m_V \sim 6.6$. From this, they derived an upper limit to the transient event rate of $0.8\ \deg^{-2}\ {\rm day^{-1}}$.
\par
Previous optical observations with high temporal resolution reported that glints of sunlight reflected from artificial satellites and space debris are significant contaminant in searching for short-duration transients, most of which are located in low Earth orbit (LEO; altitude of $\leq 2,000$ km) and geostationary orbit (GEO; $\sim 36,000$ km above the equator) (e.g., \cite{Corbett2020,Nir2021a,Nir2021b,Tanaka2025}). 
\citet{Richmond2020} and the following \citet{Arimatsu2021} conducted searches for "optical flashes" including expected optical emission from FRBs by monitoring video imaging with wide-field CMOS imagers to explore timescales of $1.5-11.5$ s and $0.1 - 1.3$ s, respectively. Their transient search programs subsequently focused on the Earth's shadow region at GEO, where contamination from sunlight glints by man-made objects is minimized. Their non-detection results placed upper limits on the event rates of optical transients of $R(\tau = 1.5-11.5\ {\rm s}) \lesssim 1.46\ \deg^{-2}\ {\rm day^{-1}}$ and $R(\tau = 0.1-1.3\ {\rm s}) \lesssim 9.1\ \deg^{-2}\ {\rm day^{-1}}$, respectively. 
\par
In this work, we conduct a systematic search for transients lasting as short as one second using consecutive 1-frame per second (1 fps) video observations targeting the Earth’s shadow region by Tomo-e Gozen, which is described in Section \ref{sec:obs}. In Section \ref{sec:ana}, we describe our transient-detection pipeline that identifies point sources with durations as short as one second in a series of 1 fps data cubes. We present the discovery of a flash-like transient candidate, TMG20200322, with a duration of $\lesssim$ 2 s in Section \ref{sec:result}, and we discuss the possible origin of this transient in Section \ref{sec:discussion}. The derived event rate of second-timescale optical transinets by comparing previous studies is also described. 

\section{Observations}\label{sec:obs}
We carry out observarions using the Tomo-e Gozen camera mounted on the 105~cm Schmidt telescope at the Kiso Observatory in Japan. Tomo-e Gozen is a wide-field mosaic CMOS camera that consists of 84 CMOS image sensors \citep{Sako2018}. Each CMOS sensor has 2000 $\times$ 1128 photosensitive pixels of size 19 $\mu$m on a side with a pixel scale of 1.189 arcsec/pixel, which corresponds to a total FoV of $20.8\ {\deg^2}$ ($39'.7 \times 22'.4$ for each sensor). Tomo-e Gozen is capable of video imaging at up to 2 fps in full-frame readout mode. In normal operation, no optical filters are used (only a glass window is attached). The wavelength coverage of Tomo-e Gozen spans approximately $370-730$ nm, with a peak quantum efficiency of 0.68 at $\lambda = 500$ nm \citep{Kojima2018}. 
\par
\begin{figure}
    \begin{center}
    \includegraphics[width=1.0\linewidth]{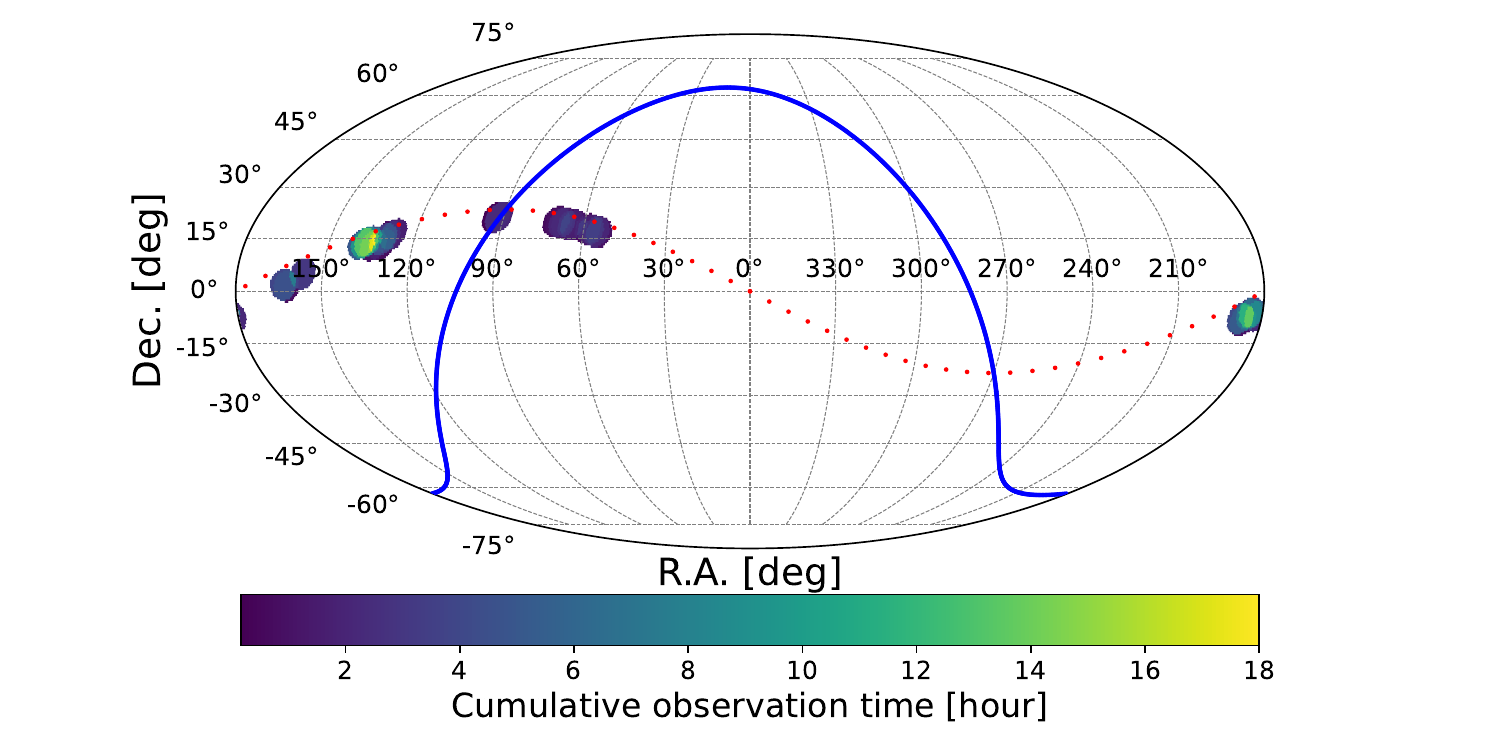}
    \end{center}
    \caption{Survey footprint of our 1 fps video observations targeting Earth’s shadow regions at GEO. Each pointing area is shown as a circle with a diameter of $\phi = 9 \deg$, corresponding to the FoV of Tomo-e Gozen. The cumulative observation time in hours is color-coded. The blue solid curve denotes the Galactic plane, while the red dotted curve marks the Ecliptic plane.
    {Alt text: All-sky map in equatorial coordinates showing the coverage of our survey.}
    }
    \label{fig:footprint}
\end{figure}

 
Observations are conducted on 28 nights between 20 November 2019 and 25 March 2020. Semi-automatic observations are performed during lunar ages 0–9 and 21–29.5 days, when the Earth's shadow region is visible at elevations $\gtrsim 30 \deg$. This lunar phase constraint is geometrically necessary, as the Earth's shadow (antisolar) direction coincides with the Moon's position during full moon. We operate Tomo-e Gozen in 1 fps mode (with an exposure time of 1.0 s), conducting up to 30 consecutive pointings (approximately one hour) at fixed sky coordinates. To ensure the FoV of Tomo-e Gozen ($\phi \sim 9 \deg$) remained within the Earth’s shadow at GEO throughout the night, we apply telescope dithering to the center of the Earth’s shadow approximately once per hour. Each video data cube generated from all 84 CMOS sensors of Tomo-e Gozen consists of 120 frames per sensor and has a data size of approximately 1.1 GB. In total, we obtain approximately 50 hours of 1-fps video data cubes with data size of $\sim 135$ TB. This translates to an areal exposure — the product of the total observed sky area and total observation time — of approximately $790\ \mathrm{deg}^2 \cdot \mathrm{hour}$\footnote{Note that the areal exposure is calculated after removing the shallow-depth data ($m_{\rm lim} > 16$ mag) and masked pixels in our transient search. See Section \ref{sec:ana} for more details.}, which is about 16 times greater than the previous work by \citet{Richmond2020} which used the prototype model of Tomo-e Gozen ($\sim 50\ \deg^2 \cdot {\rm hour}$). 
Figure~\ref{fig:footprint} shows the footprint of our Earth’s shadow observations. Overlapping regions with longer exposure times are visible, resulting from shifting the pointing center. 
%%%
%%%
\par
\begin{figure}
    \begin{center}
    \includegraphics[width=1.0\linewidth]{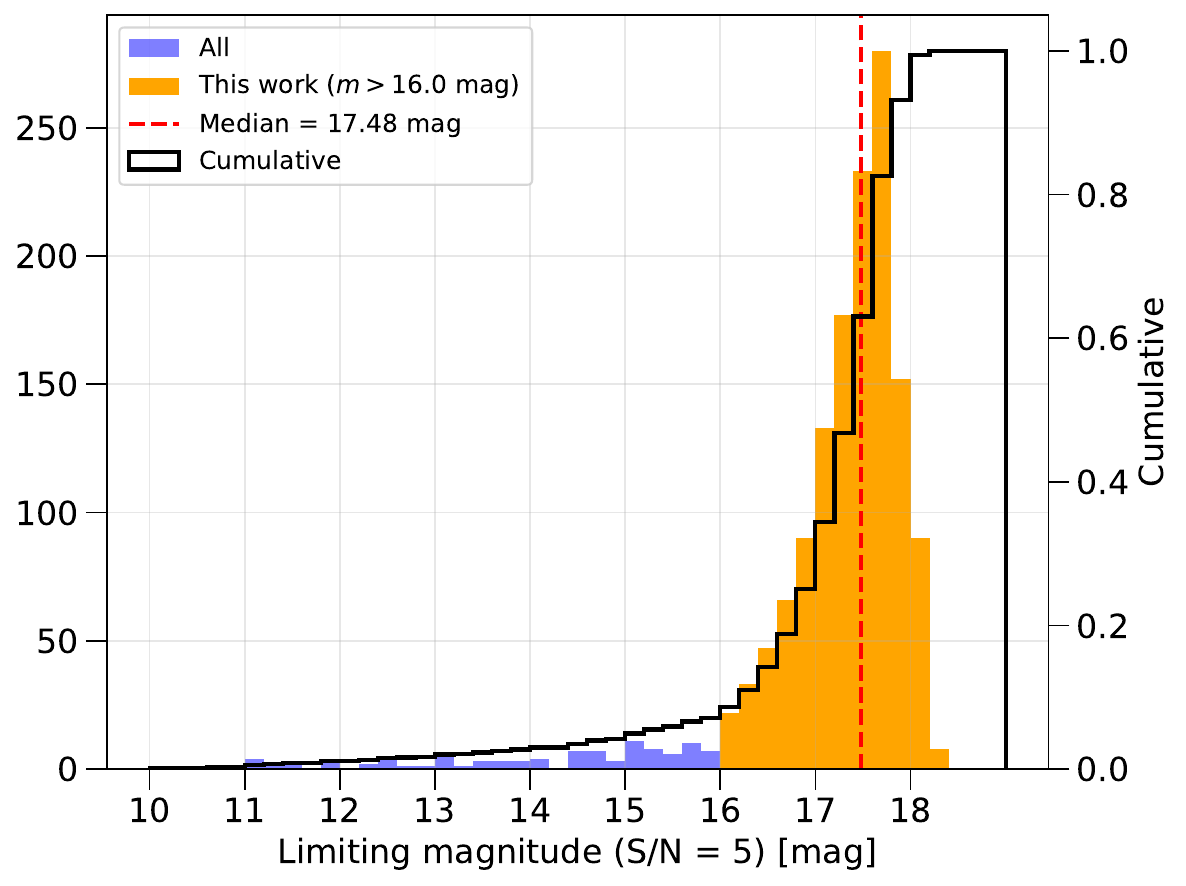}
    \end{center}
    \caption{Histogram of the limiting magnitudes for our 1 fps video observations. The limiting magnitude is derived from each data cube consisting of 120 frames, using CMOS sensor ID 133, which has median sensitivity among the 84 sensors. The red vertical line indicates the median limiting magnitude ($\rm S/N = 5$), which is 17.48 mag. The black curve represents the cumulative distribution of the histogram, corresponding to the right vertical axis. For the transient search analysis described below, we set a limiting magnitude cut at $m_{\rm lim} = 16.0$ mag and exclude frames with shallower limiting magnitudes. The binning interval is 0.2 mag.
    {Alt text: A graph of histograms with a reference line.}
    } 
    \label{fig:limmag}
\end{figure}

Raw images obtained by the Tomo-e Gozen camera are
automatically reduced via the Tomo-e Gozen analysis servers \citep{Sako2018},
where basic reduction processing of dark frame subtraction and flat-field correction are performed.  For the reduced images, astrometric calibration is performed employing the $Gaia$ DR2 catalog (epoch:
2015.5) \citep{GaiaCollaboration2018} using Astrometry.net software \citep{Lang2010}. We use the reduced and astrometrically calibrated frame images in our analysis. 
We derive the photometric zero points of the Tomo-e Gozen images by using the $Gaia$ DR2 $G$-band photometric system \citep{GaiaCollaboration2018}. We calculate the photometric zero point by cross-matching sources on each frame image of Tomo-e Gozen with $Gaia$ DR2 catalog sources whose magnitude ranges $12 \leq m_G \leq 16$ mag using the VizieR database service \citep{Ochsenbein2000}. Color correction is not performed in our analysis. We omit airmass correction because we perform relative photometry for sources within the same image. Figure \ref{fig:limmag} presents a histogram of the limiting magnitudes of our video frame images. We exclude frame images with limiting magnitudes shallower than $m_G = 16$ mag for the following transient search. The median value is 17.48 mag (signal-to-noise ratio $\rm S/N = 5$), which we adopt as the limiting magnitude of this work. 
%%%
%%%

\section{Search for second-timescale transients}\label{sec:ana}
\subsection{Transient detection}
\begin{figure*}
    \begin{center}
    \includegraphics[width=1.0\linewidth]{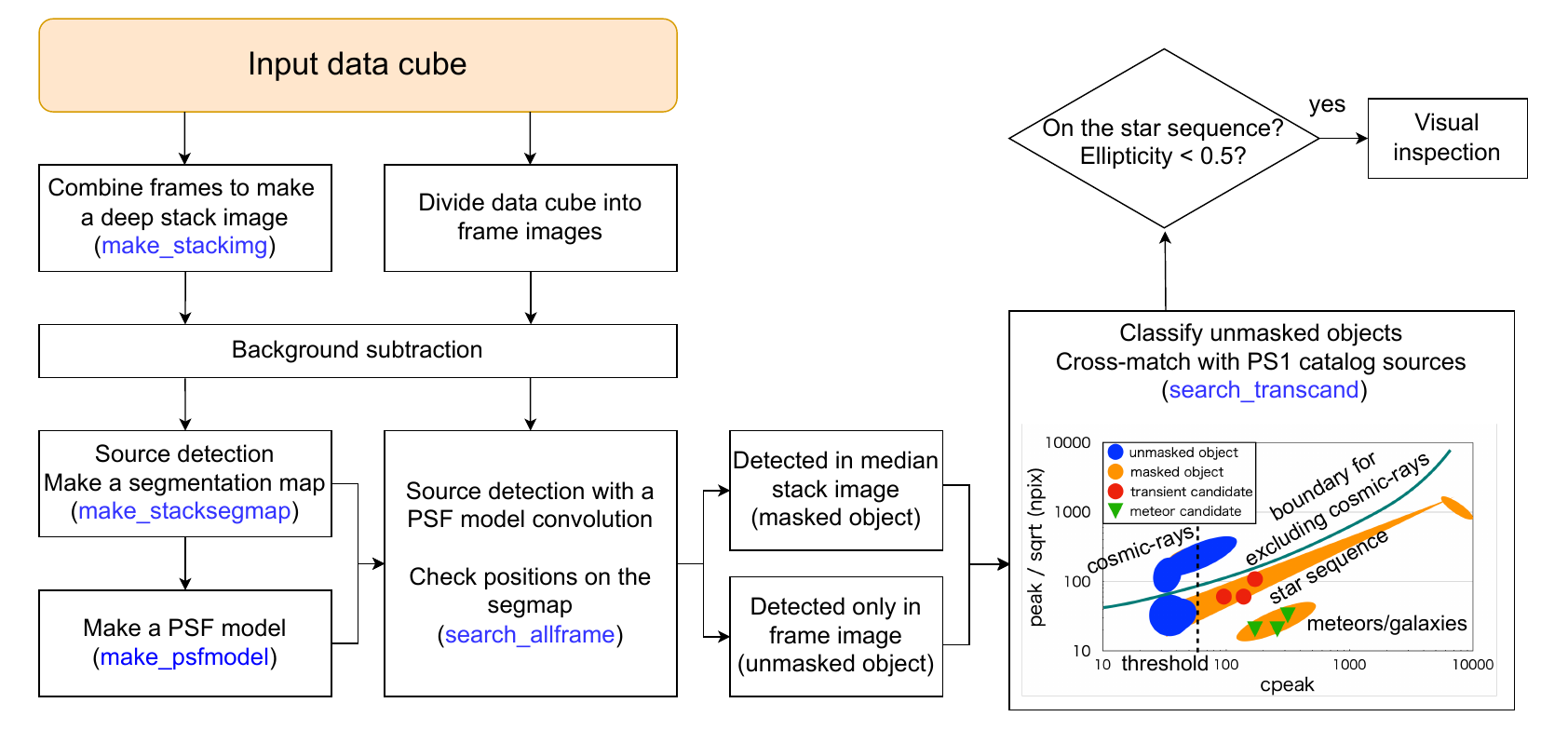}
    \end{center}
    \caption{Flowchart of our transient detection pipeline. The input data cube is processed through two analytical flows: the stacking process (core functions are make\_stackimg, make\_stacksegmap and make\_psfmodel shown as blue in the brackets) and per-frame processing (search\_allframe). Objects detected in individual frames are compared against a binary image called a segmentation map (segmap), which contains positional information of detected objects in the stacked image. Unmasked objects are those that are detected only in some frames but are not detected in the stacked image. These unmasked objects can be transient candidates, which are filtered based on their PSF profile and ellipticity parameters to remove false positives (e.g., cosmic rays) and meteors. In the right-hand figure, the schematic diagram illustrates the classification method for unmasked objects (search\_transcand process). Unmasked objects located in the blue regions are classified as false positives, whereas those appearing as red circles along the stellar sequence are identified as transient candidates.} After passing these selection criteria, human visual inspection of the candidate images is performed. {Alt text: A flowchart showing the sequence from data input through processing to transient candidate evaluation.}
    \label{fig:flowchart}
\end{figure*}

To search for second-timescale transients from video data, we develope a custom transient-detection pipeline. The flowchart of our pipeline is shown in Figure \ref{fig:flowchart}. 
In the left part of the flowchart, the input data cube consisting of 120 frames is stacked along the time axis to generate a single, deep image. The stacked image serves as a reference for stationary, non-transient objects. We adopt median stacking to ensure that transient events appearing in only a few frames are excluded from the stacked image. Our pipeline is capable of detecting transient events with durations as short as one second from the 120-frame data cube (See Section \ref{sec:discussion} for more details). The stacked images are then background-subtracted and source detection is performed using the \texttt{SEP} package \citep{Barbary2016}, a Python wrapper of \texttt{SExtractor} \citep{Bertin1996}. At the same time, a segmentation map of the same size as the stacked image, hereafter referred to as ``segmap'', is generated. In the segmap, the pixel coordinates of the detected objects are recorded along with their corresponding identification numbers\footnote{A segmentation map can optionally be generated during the source detection process with \texttt{SEP}.}. For each stacked image, an average point spread function (PSF) is modeled using bright stars by fitting a 2D Gaussian function and incorporating the residuals of the fit (observed brightness of the stellar profile) using \texttt{PythonPhot} \citep{Jones2015}, a Python wrapper of the core functions of \texttt{DAOPHOT} \citep{Stetson1987}. 
\par
\begin{figure}
    \begin{center}
    \includegraphics[width=1.0\linewidth]{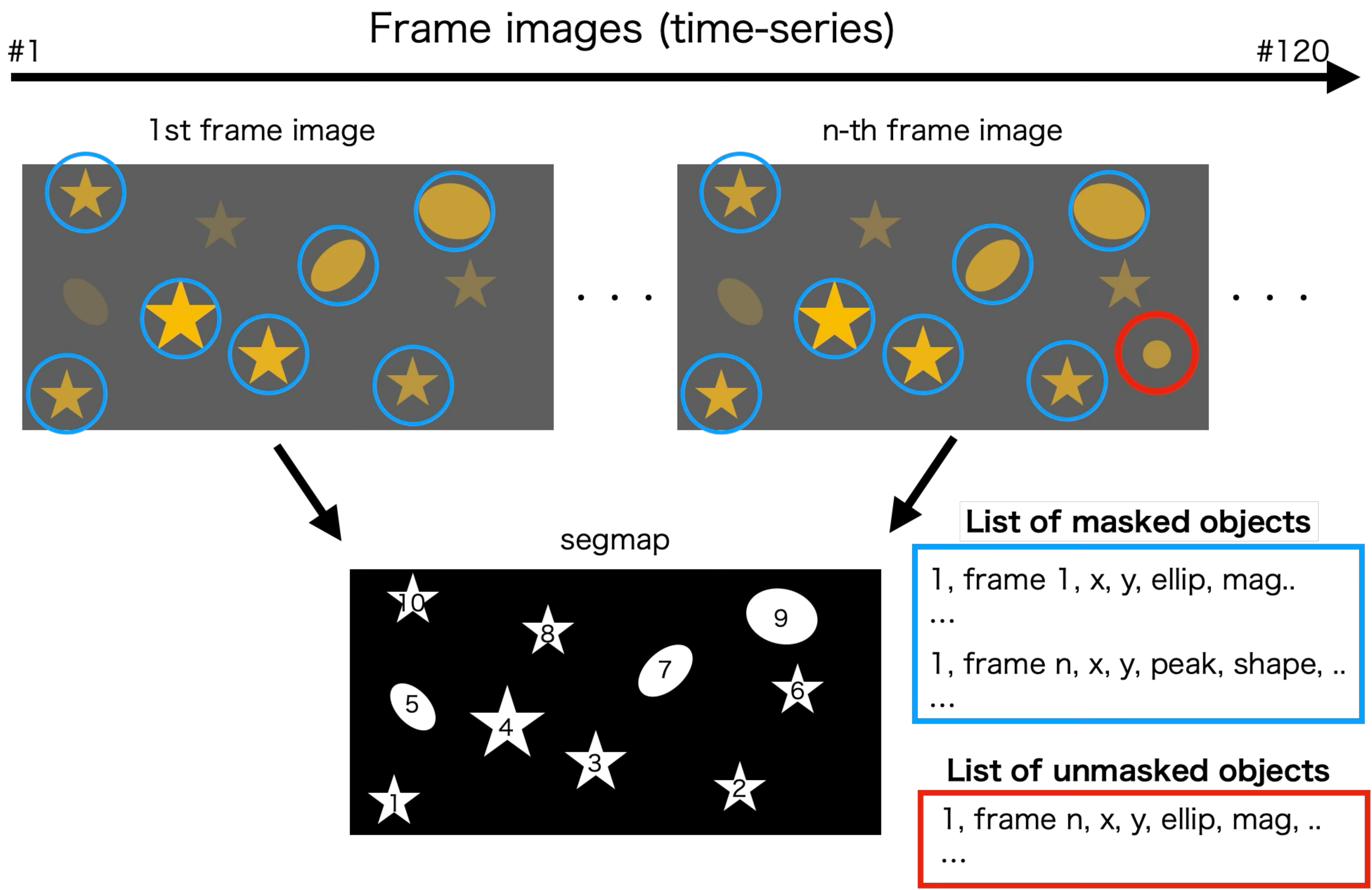}
    \end{center}
    \caption{A schematic picture of how to assign masked/unmasked objects with a segmap. Each object detected in each frame (marked as red and blue circles) is assigned segmap IDs by referring to the segmap. Finally two output catalogs of masked objects and unmasked objects including transient candidates are created. We then search for transients from the unmasked objects. 
    {Alt text: Diagram showing how objects detected in time-series frames are compared with a segmentation map and sorted into masked and unmasked object lists.}
    } 
    \label{fig:transient_detection}
\end{figure}

In the right part of the flowchart, source detection is performed on each background-subtracted frame by convolving with the PSF model generated from the stacked image with the detection threshold of $1\sigma$ of the sky-background noise. Aperture photometry is then carried out for each detected object using two fixed aperture radii of 3 and 5 pixels (approximately 3.6 and 6.0 arcseconds, respectively). Sources detected in each frame image are classified into two categories based on the corresponding segmap: 
%%%
\begin{enumerate}
    \renewcommand{\theenumi}{\roman{enumi}}
    
    \item masked objects, which are detected in the median stack image
    \item unmasked objects, which are only detected in frame images
\end{enumerate}
%%% 
Figure \ref{fig:transient_detection} illustrates how to search transient events in our pipeline. 
Objects with duplicate segmap IDs are excluded during this process. We only use the unmasked objects for the following transient selection. This reduces the total areal exposure of our transient search, which is taken into account for the event rate calculation in Section \ref{sec:discussion}. 
%%%
%%%
%%%
\par
\subsection{Selection criteria}
To reduce false positives from unmasked objects, we employ a classification diagram based on the parameter ${\rm peak/\sqrt{npix}}$, ratio of the peak count and square root of the number of pixels (npix) and the peak count derived on the model-PSF-convolved image (cpeak). The use of using these parameters is based on the previous work aimed at classfying stars and galaxies \citep{yamagata1986a}. This diagram reflects the brightness profile of astronomical objects and is effective in distinguishing between stars and non-stellar objects. Objects with large peak and small npix values -- such as cosmic ray hits and electrical noise -- are located above the star sequence in Figure \ref{fig:flowchart}. For extended objects, such as galaxies (masked objects) and meteors (unmasked objects) with large npix and moderate peak values are located under the star sequence. We use an empirically derived boundary to select point sources from unmasked objects as shown in Figure \ref{fig:flowchart}. We further limit unmasked objects with  ellipticity $e<0.5$ defined as $e=1-b/a$, where $a$ and $b$ are the semi-major and semi-minor axes of an object. Transient candidates that meet the selection criteria described above are finally visually inspected. We also cross-match with Pan-STARRS DR1 sources via the VizieR database service \citep{Ochsenbein2000} using a search radius of $r_{\rm PS1} = 2.0$ arcseconds to identify potential optical counterparts. 

\section{Results}\label{sec:result}
\subsection{Source classification}
In the analysis described above, 392 events pass our selection criteria. 
We classify candidate sources into several categories based on the shape, time variations of positions and brightness on the subsequent frame images. The categories we used for source classification are as follows. 
\begin{enumerate}
    \item Transient candidate
    \item Moving object
    \item Meteor
    \item Electronical defect
    \item Cosmic ray
    \item transient caused by clouds
    \item Image deviation
    \item Positional offset of stars
    \item Bad pixel (including hot/cold pixel)
    \item Spider pattern by bright stars
\end{enumerate}
Moving objects include asteroids in the solar system and may also
include artificial satellites or space debris orbiting outside GEO in Earth's shadow observations. We separate the category of moving objects and meteors in terms of their apparent shapes, i.e. point sources and elongated ones. The other categories of (4 - 10) are false positives caused by external factors or instrumental problems. We visually inspect the moving images of each candidate event and find one possible transient candidate. 
\subsection{A candidate of second-timescale transients}
\begin{figure}
    \begin{center}
    \includegraphics[width=1.0\linewidth]{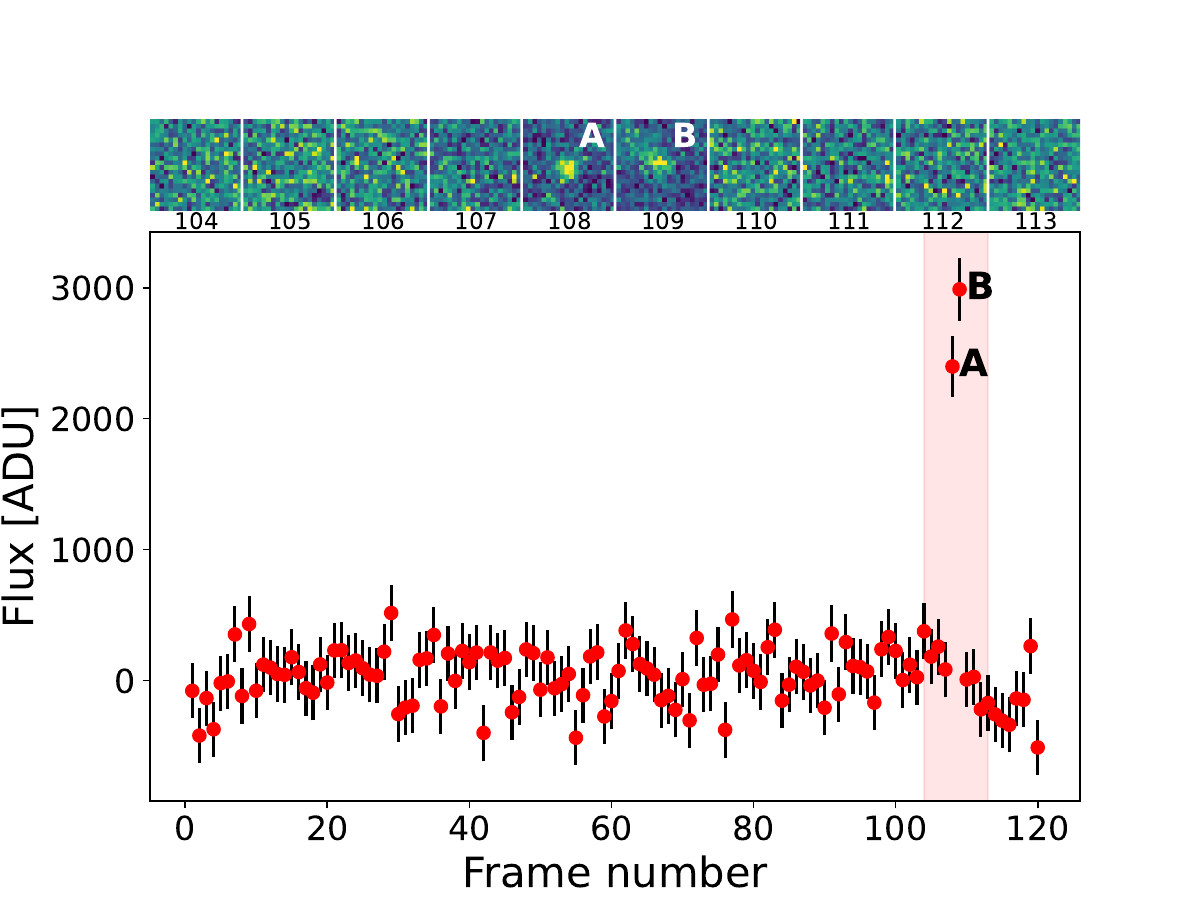}    
    \end{center}
    \caption{Forced photometry of the transient candidate TMG20200322. The error bars indicate the 1$\sigma$ uncertainties calculated from the sum of the Poisson noise and the background errors. We highlight the red-shaded region at the detection frames of TMG20200322 (frame number 108 and 109 labeled as A and B) and the corresponding 10 consecutive frame image cutouts (frame number from 104 to 113) are shown in the upper panel. Each cutout has 24 arcsec (20 pixels) on a side, oriented with North up and East to the left. Lower and upper percentiles of 1.0 \% and 99.5 \% of the pixel values are used to normalize the scale of each cutout.
    {Alt text: Graph showing flux versus frame number, highlighting two peaks labeled A and B. Above the graph, small image cutouts from consecutive frames are displayed.}
    } 
    \label{fig:lightcurve}
\end{figure}

\begin{figure}
    \begin{center}
    \includegraphics[width=1.0\linewidth]{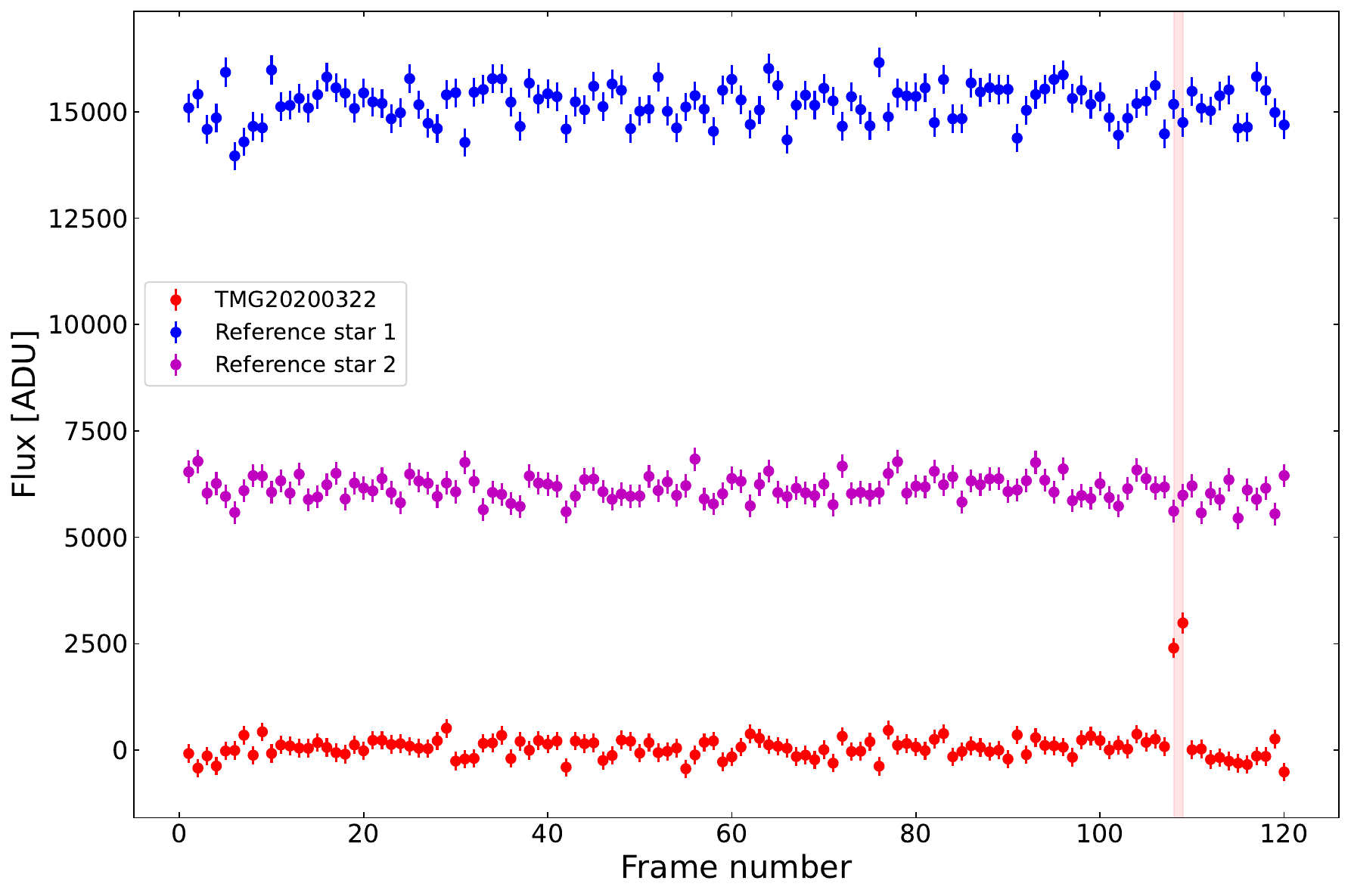}
    \end{center}
    \caption{Comparison of light curves of the two reference stars selected from the $Gaia$ DR3 catalog close to the position of TMG20200322. Ref. star 1; $G = 14.96$ mag, (RA, Dec)$_{\rm J2000.0}$ = (\timeform{12h18m32.0s}, \timeform{-4D55'58.5''}), and Ref. star 2; $G = 15.95$ mag, (RA, Dec)$_{\rm J2000.0}$ = (\timeform{12h18m14.0s}, \timeform{-4D54'34.9''}). Red vertical shaded region indicates the time of detection of TMG20200322 (108-th and 109-th frames). The error bars denote 1$\sigma$ photometric uncertainties.
    {Alt text: Light curves of a transient candidate and two reference stars with detection highlighted.}
    }
    \label{fig:lc_comparison}
\end{figure}

The transient candidate TMG20200322 is detected on the image taken on 22 March 2020 at 15:02:25.043 UT. The discovery coordinates are (RA, Dec)$_{\rm J2000.0}$ = (\timeform{12h18m31.6s}, \timeform{-4D55'07.5''}) and its Galactic coordinates are $(l, b)$ =
$(287.76391, 56.98059)$ in degrees. 
% ra, dec = 184.63183, -4.91875 in degrees
We present a light curve derived from forced photometry at the coordinates of TMG20200322 in Figure \ref{fig:lightcurve}. TMG20200322 has a $\rm S/N \sim 10.2$ and $\sim 12.4$ for the 1st and 2nd frames measured by aperture photometry with a radius of 6 arcsec. The corresponding $Gaia\ G$-band magnitude is $m_G$ = 16.89 $\pm$ 0.11 mag and 16.65 $\pm$ 0.09 mag for the first and second detection frames, respectively (Table \ref{tab:TMG20200322}). We compare the light curve of TMG20200322 with those of the field stars on the same image in Figure \ref{fig:lc_comparison}. The light curves of reference stars show no systematic variations in frames 108 and 109, confirming that the observed brightening is intrinsic to the transient source. 
%%%
%%%
\begin{table*}
    \caption{Observed properties of the transient candidate TMG20200322 discovered on March 22, 2020. The UT time listed in the table represents the exposure interval, indicating the start and end times of integration for the detection frames. The $1 \sigma$ uncertainty of $Gaia$ $G$-band magnitude ($m_G$) and ellipticity is indicated in the brackets. $x$ and $y$ are the position of the flux barycenter and $x$peak and $y$peak are the position of the peak pixel. The ellipticity and source position measurements were performed with a threshold of $1\sigma$ of the background rms.}
    \centering
    \begin{tabular}{lcccccc}
          \hline \hline Frame & UT (exposure interval) & $m_G$ & Ellipticity & $x,\ y$ & $x$peak, $y$peak \\ No. & & & & (pixel) & (pixel)\\ \hline 
          108 & 15:02:24.043 $-$ 15:02:25.043 & 16.89 (0.11) & 0.16 (0.13) & 1603.0, 1008.0 & 1603.0, 1008.0 \\ 109 & 15:02:25.043 $-$ 15:02:26.043 & 16.65 (0.09) & 0.50 (0.05) & 1605.5, 1005.2 & 1603.0, 1007.0 \\ \hline
    \end{tabular}
    \label{tab:TMG20200322}
\end{table*}
%%%
\par
\begin{figure*}
    \begin{center}
    \includegraphics[width=1.0\linewidth]{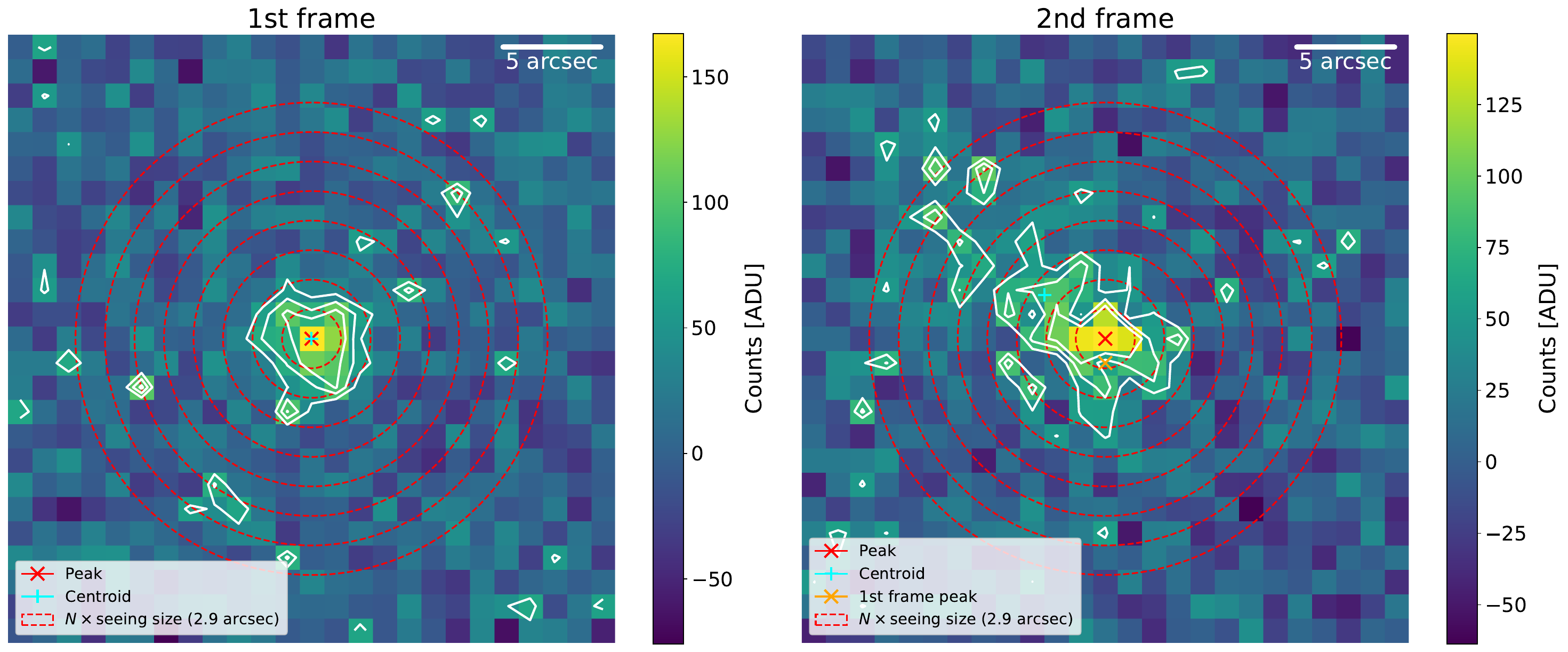}
    \end{center}
    \caption{Zoomed-in images of TMG20200322. The contours with white lines at 2.0, 3.0, and 4.0 $\sigma$ levels of the background noise are overlaid. Peak (red) and centroid (cyan) pixels are marked in each image (1st frame peak pixel is also marked as orange in the 2nd frame image). The red dashed circles indicate $N$ times the size of seeing FWHM (2.9 arcsec). The cutout image size is $30'' \times 30''$ and the top is north and the left is east.} 
    {Alt text: Two zoomed-in frame images with contour lines and overlaid red dashed circles.
    }
    \label{fig:cutout_img}
\end{figure*}
\begin{figure*}
    \begin{center}
    \includegraphics[width=1.0\linewidth]{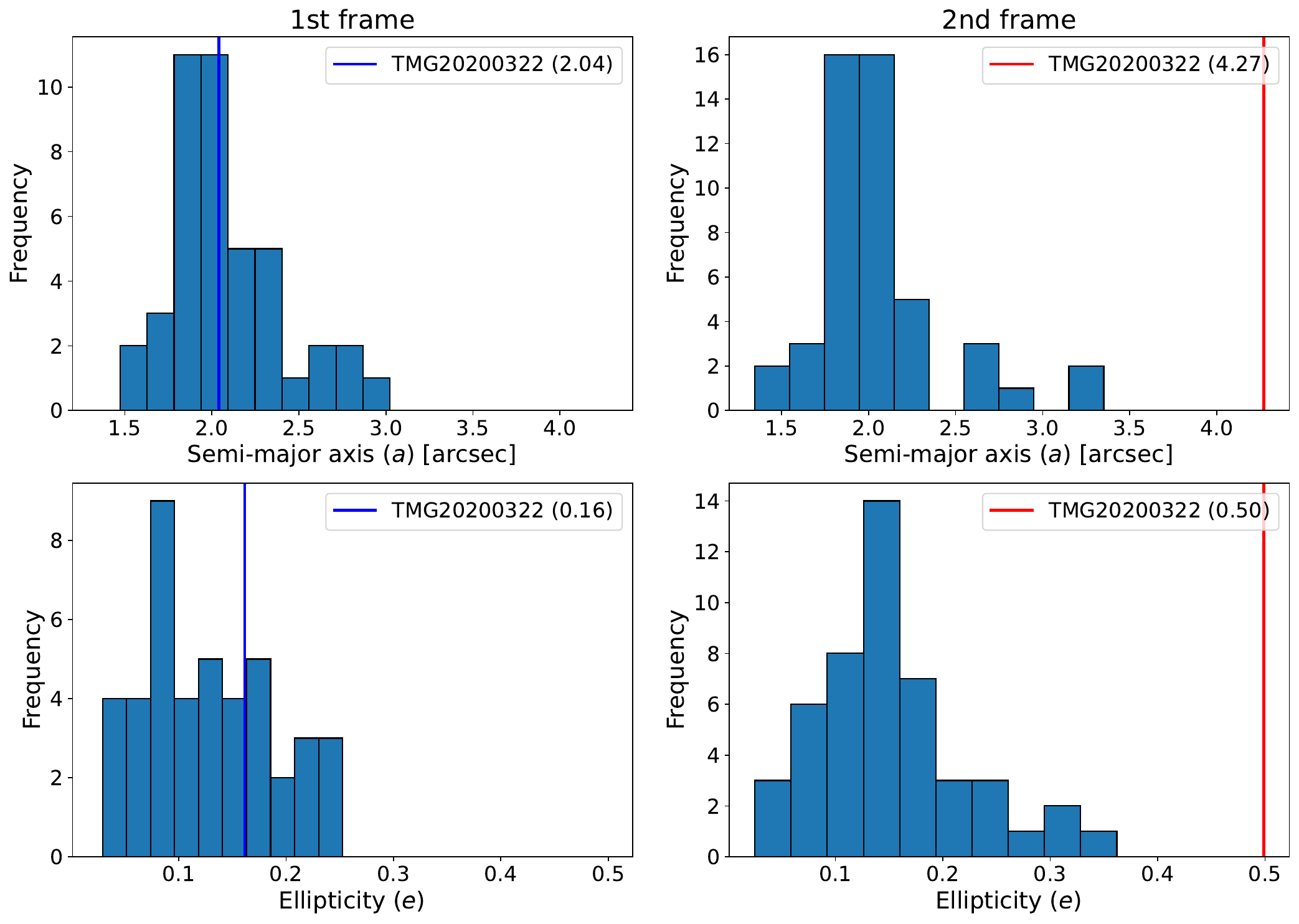}
    \end{center}
    \caption{
    Distribution of shape parameters for field stars compared with TMG20200322. Top panels: semi-major axis (a) in arcseconds; bottom panels: ellipticity (e). Left and right panels correspond to the 1st and 2nd detection frames, respectively. Vertical lines show the measured values for TMG20200322 (blue: 1st frame; red: 2nd frame), as indicated in the brackets in the figure legends. Field stars have S/N ratios between 10 and 15. All shape parameters are measured at an isophotal level of $1\sigma$ above the background noise. 
    {Alt text: Four histogram panels arranged in a 2-by-2 grid.}
    }
    \label{fig:ellip_hist}
\end{figure*}

In Figure \ref{fig:cutout_img}, we show zoomed-in image cutouts of the two detection frames of TMG20200322. We find a variation in the spatial brightness distribution between the two frames. Specifically, the second frame image exhibits a directional elongation toward the northeast. The semi-major axis measured at an isophotal level of $1 \sigma$ of the background rms is 4.27 arcsec, corresponding to a significance of $6.1 \sigma$ relative to the distribution of the field stars (Fig. \ref{fig:ellip_hist}). TMG20200322 also exhibits significantly higher ellipticity in the second frame ($e = 0.50 \pm 0.05$) compared to those of the field stars, although the first frame has typcal value ($e = 0.16 \pm 0.13$). 
The flux barycenter between the two frames
differs by $\sim 4.5$ arcsec, which is significant given the seeing condition
($\sim 3$ arcsec) and the pixel scale of the Tomo-e Gozen camera ($1.189$ arcsec/pixel). On the other hand, the peak pixel positions in the two frames are nearly identical, which differs by just one pixel (see Table \ref{tab:TMG20200322}). This suggests that TMG20200322 could be a stationary object showing an extended component in its brightness distribution.
\par
%%%
%%%
%%%
%%% 
\begin{figure}
    \begin{center}
    \includegraphics[width=1.0\linewidth]{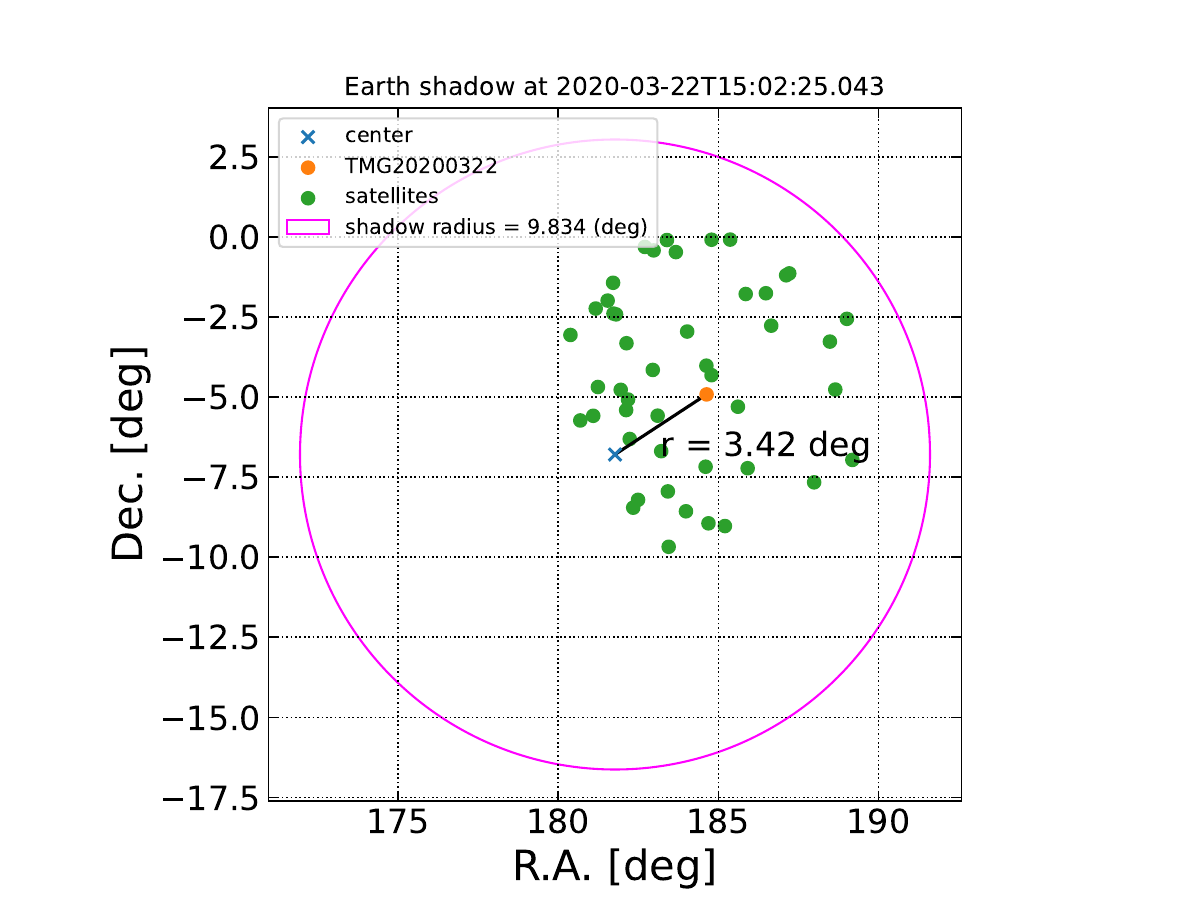}
    \end{center}
    \caption{The coordinates of TMG20200322 and the Earth's shadow region at GEO. The blue cross indicates the center of the Earth's shadow at the time of discovery of TMG20200322, and the shadow region is represented by the magenta circle. TMG20200322, marked as an orange point, is located inside the shadow, approximately $3.42 \deg$ away from the shadow center. The green points represent artificial bodies found within a 5-degree radius from the coordinate of TMG20200322 by the Space Track database.
    {Alt text: Scatter plot in equatorial coordinates showing an orange point inside a magenta circle, with a blue cross marking the center and multiple green points scattered around.}
    }
    \label{fig:TMG20200322_coo}
\end{figure}
%%%
Since our observations target the Earth's shadow region at GEO, reflections from artificial objects within the shadow are unlikely to be detected. However, the possibility of reflections from objects located outside the GEO region cannot be entirely excluded. We obtain a list
of cataloged artificial objects within a radius of 5 degrees of the coordinates of TMG20200322 at the time of discovery using the Space Track database\footnote{https://www.Space Track.org/}. The coordinates of TMG20200322 and the Earth's shadow region at the time of discovery are shown in Figure \ref{fig:TMG20200322_coo}. For TMG20200322 to move out of the Earth's shadow region, its physical distance would need to be about 3 times that of GEO ($\gtrsim 100,000$ km). The most distant cataloged object ("CZ-3B DEB") is $46,427$ km from the Earth's surface. Therefore, it is unlikely that TMG20200322 is a cataloged artificial object. If we assume that TMG20200322 is an object illuminated by reflected sunlight at 100,000 km from Earth, its characteristic diameter would range from 0.5 to 1.4 m depending on the albedo values assumed (0.1 and 0.85). However, the physical size estimated from the elongation in the second frame is $\sim 4.5$ km, which contradicts the size derived from the observed magnitude. This discrepancy indicates that the assumption that TMG20200322 is an artificial object reflecting sunlight is unlikely. 
\par
%---Host galaxy---%
No host galaxy candidates or point source counterparts for TMG20200322 are detected in the Tomo-e Gozen stacked image. We investigate the Pan-STARRS1 (PS1) 3$\pi$ survey source catalog (limiting magnitude of $m_{lim} \sim 23$ mag in the 
$r_{\rm PS1}-$ band) \citep{Chambers2016} and found no possible optical counterparts within a radius of 5 arcsec. We use the archival Pan-STARRS1 (PS1) stack images in $g,r,i,y,z$-bands to constrain the magnitude of the optical candidate. We calculate the $5 \sigma$ limiting magnitude of the PS1 images by putting
circular apertures with radius of 5 pixels ($\simeq 1.25''$) at random sky locations. We derive limiting magnitudes of $23.25,\ 22.45,\ 22.83,\ 21.22,\ 22.26$ mag for the $g,r,i,y,z$ bands, respectively. 
%---Transients---%
No transient event including GRBs within a radius of 1 arcmin of the TMG20200322 coordinates has been reported on the Transient Name Server
(TNS)\footnote{https://www.wis-tns.org} before or after the time of its discovery. We searched the following catalogs of multiwavelength survey sources: Two Micron All Sky Survey (2MASS; near-IR), Galaxy Evolution Explorer (GALEX; UV) and Fermi Gamma-ray Space Telescope via NASA's HEASARC archive catalog service\footnote{https://heasarc.gsfc.nasa.gov}, and the Very Large Array/FIRST survey (1.4 GHz Radio)\footnote{http://sundog.stsci.edu/cgi-bin/searchfirst}. 
No counterpart is found within a 10 arcsec radius of the transient. 

\section{Discussions}\label{sec:discussion}
\subsection{Near-field effects}
We first discuss the possibility that the transient candidate TMG20200322 could be explained by near-field effects: objects subject to temporary local illumination (e.g. aircraft, dust, insects), and optical artifacts such as ghost images or stray light. 
\par
For nearby illuminated objects, the Kiso Schmidt telescope (diameter of $105$~cm) focused at infinity would produce significantly defocused images. For example, a point source at a distance of 10~km, corresponding to a typical altitude of an aircraft, would be defocused by approximately 20~arcseconds. Objects at closer distances would be even more severely defocused. This is inconsistent with the observed sharp PSF ($\sim$3~arcsec FWHM), making near-field explanations increasingly implausible. Moreover, an aircraft would be expected to appear bright and move across the image during the exposure, producing a continuous trail rather than the discrete elongated PSF observed in a single frame.
\par
Regarding optical artifacts such as ghost images or stray light, we note the following: (1) Ghosts do not vary on timescales of a single exposure of 1~s. (2) While stray light from external sources could enter the optical system by chance and form an image, such artifacts generally do not produce images as sharp as observed. We have found no similar artifacts that could be attributed to systematic optical effects. These considerations allow us to confidently rule out near-field effects, ghost images, and stray light as explanations for TMG20200322.
%%%
\subsection{PSF distortion due to atmospheric fluctuations}
Atmospheric fluctuations blur the light from stars, distort the PSF in the image. \citet{MegiasHomar2023} investigated how exposure time affects the PSF of very short-duration Fast Optical Bursts (FOBs), which are assumed to be the optical counterparts to FRBs. They simulated observations of the Rubin/LSST to compare spatial brightness distributions of FOBs, assuming a FOB duration of $15$ ms. The results showed that FOBs with 15 ms duration exhibit a PSF that is distinctively different from that of steady sources when observed with a LSST's nominal exposure of 15 s. However, the spatial extent of the atmosperically distorted PSF remains confined within the seeing disk. 
%%%
\subsubsection{Comparison with field stars}
\begin{figure*}
    \begin{center}
    \includegraphics[width=0.9\linewidth]{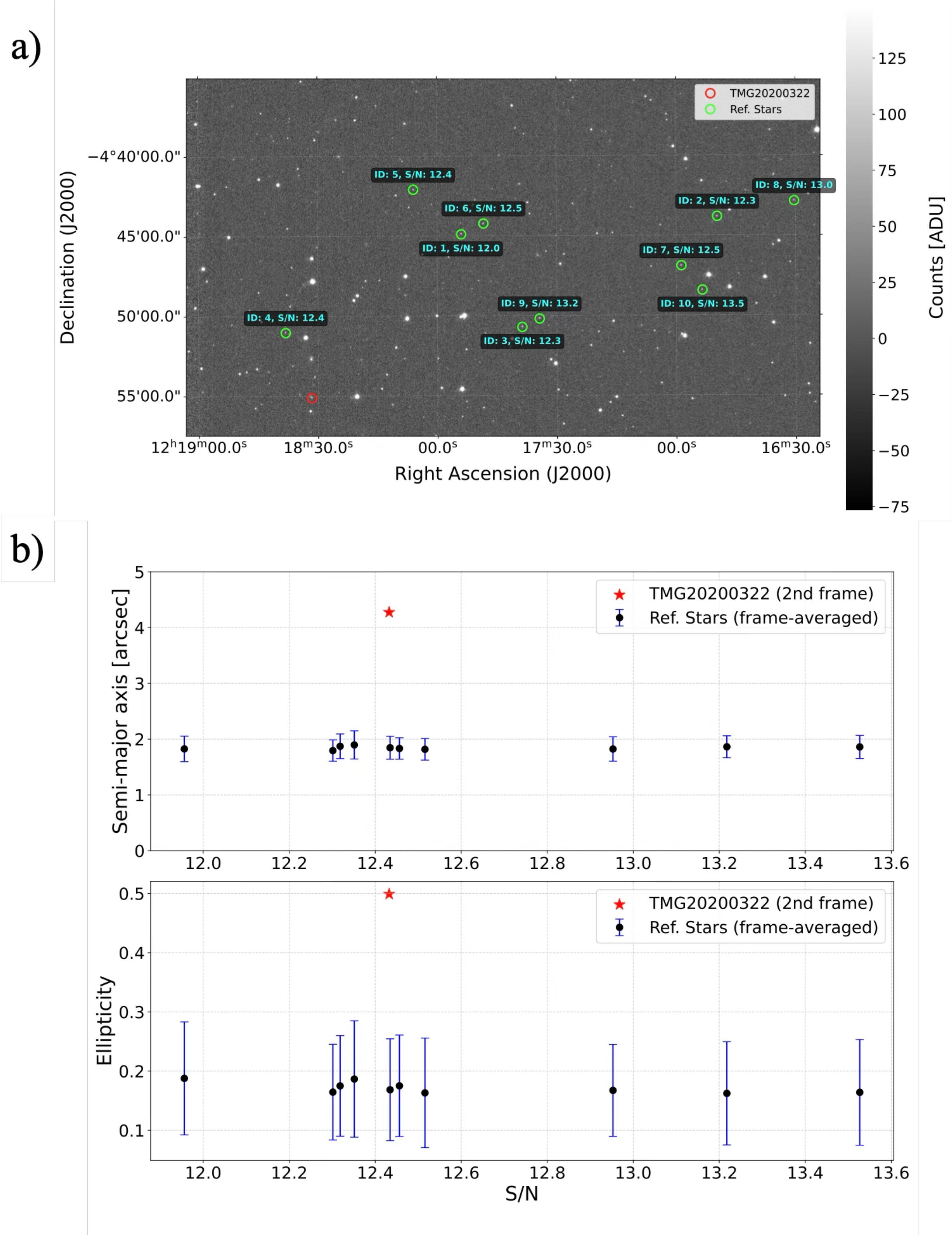}
    \end{center}
    \caption{Shape variability of the field stars in the detection data cube of TMG20200322. (a) Frame image (2nd detection frame of TMG20200322) showing the positions of the 10 field stars, labeled with their source IDs, S/N values, and the number of detected frames. (b) Semi-major axis and ellipticity of the field stars as a function of S/N, with the values for TMG20200322 indicated by red stars. Error bars for each plot indicate $1 \sigma$ uncertainty in both measurements.
    {Alt text: Two-panel figure: the left panel shows field star positions marked on the image, while the right panel presents two measurement plots, with TMG20200322 highlighted as red star.} 
    }
    \label{fig:field_star}
\end{figure*}

%%%
To verify that the elongated image of TMG20200322 is not due to 
atmospheric fluctuations, we perform a shape variability comparison with field stars. We select 10 reference stars within the same 120-frame data cube with signal-to-noise ratio (S/N) ranging $10-15$ from the $Gaia$ DR3 catalog \citep{GaiaCollaboration2023}, ensuring comparable photometric quality to that of TMG20200322 (S/N of $\sim$10.2 and $\sim$12.4 on the 1st and 2nd frames). To select single stars with good astrometric solutions, we impose the following selection criteria on the $Gaia$ sources: $\texttt{non\_single\_star} = 0$ and renormalised unit weight error $\texttt{ruwe} < 1.4$. 
We measure semi-major axis and ellipticity for each source using \texttt{sep} with a threshold of $1\sigma$ of the background rms, in the same manner as was done for TMG20200322. 
%%%
Figure \ref{fig:field_star} shows the positions of the field stars used in this analysis. It also plots their measured semi-major axes and ellipticities against their S/N. 
Compared with the field stars, TMG20200322 exhibits significantly larger values for both shape parameters. The mean semi-major axis ($a$) and ellipticity ($e$) of the field stars are $\langle a \rangle = 1.85 \pm 0.22$ and $\langle e \rangle = 0.17 \pm 0.10$, respectively. This low ellipticity is consistent with the nearly circular PSF ($e \approx 0$) expected from atmospheric seeing, confirming that the field stars are not systematically elongated. In contrast, TMG20220322 exhibits a semi-major axis of $a \approx 4.3$ and an ellipticity of $e \approx 0.5$. These values correspond to deviations of $11.1 \sigma$ and $3.2\ \sigma$ from the mean of the reference population, respectively. These results demonstrate that the elongation of TMG20200322 is not caused by atmospheric or systematic instrumental effects, but is instead of intrinsic origin. 
%%%
%%%
\subsubsection{Constraints from high-speed observations}
To place constraints on the timescale of the flash duration of TMG20200322, we perform additional high-speed observations of the Pleiades open cluster at 57 fps (17.4 ms exposure time) using the partial readout mode of Tomo-e Gozen. While these data were taken 
under different observing conditions (mean seeing size of $4.0\pm 0.3$ arcsec) than the TMG20200322 detection, they provide useful constraints on atmospheric effects on the event timescale (e.g., \cite{MegiasHomar2023}). We use stars with S/N of $10\sim15$ as in the previous analysis. 
\par
We find that TMG20200322 exhibits a significantly higher semi-major axis and ellipticity than stars observed at 57 fps, with deviations of 15.5$\sigma$ and 6.1$\sigma$, respectively. Although the atmospheric conditions for this comparison are not identical, the degree of elongation observed in TMG20200322 is unlikely to be replicated in images of stars taken with $\sim20$ ms exposure time. Therefore, despite the limitations, this analysis strongly suggests that extreme atmospheric seeing alone cannot account for the elongation of TMG20200322. 
\subsection{Potential astrophysical origins}
Here we discuss the potential astrophysical origins of TMG20200322 based on its observed properties and compare the corresponding event rates. We consider two scenarios for the source: (a) impact flash caused by a collision between two celestial bodies, and (b) a terrestrial atmospheric phenomenon, specifically a head-on meteor. Although it is difficult to explain the elongated component of the TMG20200322 image by atmospheric fluctuations alone, we also consider possible origins of TMG20200322 under the assumption that the PSF is not elongated in the Appendix. 
%%%
\subsubsection{Meteoroid Impact Flash on Near-Earth Asteroid Surfaces}
To account for the elongated component of the TMG20200322 image and its brief duration ($\lesssim 2$ s), we consider an impact flash produced by the collision of two celestial bodies. When a centimeter-sized meteoroid impacts the Moon at velocities of several tens of kilometers per second, part of its kinetic energy is converted into luminous energy, and ground-based telescopes typically observe a flash of magnitude $5 \sim 10$ 
in visible to near-infrared light 
\citep{BellotRubio1998,Ortiz2000}. This phenomenon, known as a lunar impact flash, typically lasts less than 1.0 s (e.g., \cite{Suggs2014,Liakos2024}). By replacing the Moon with a Near-Earth Asteroid (NEA), we estimate the detection rate of meteoroid impact flashes on NEA surfaces. 
\par
Although the distance to TMG20200322 is unknown, we estimate an upper limit to the event's distance based on the elongation of the image. The extended component from the core of the PSF ($2 \times$ FWHM $\sim$ 5.8 arcsec) is approximately $10$ arcsec (see Figure \ref{fig:cutout_img}). Due to the temporal resolution of our observations is one second, the actual duration of TMG20200322 may be shorter than the exposure time. We assume that the apparent image elongation is produced by ejecta moving an angular distance of 10 arcsec over a period of one second. We use the typical collision velocity between NEAs \citep{Bottke1993} of $15\ {\rm km\ s^{-1}}$ to calculate the length that the ejecta moves in one second. The ejecta velocity caused by a collision event is known to be in the range of $v_{\rm ej} \sim 10^{-4} - 10^{-1} v_{\rm col}$, where $v_{\rm col}$ is the collision velocity \citep{Ferrari2022}. Since the elongation on the image is the projection component of the ejecta on the celestial plane, we can estimate the ejecta length $l_{\rm ej} = v_{\rm ej}\cos{\theta}\cdot t$, where $\theta$ is the projection angle. Substituting maximum ejection velocity of $v_{\rm ej} = 1.5$ km ${\rm s}^{-1}$, $\theta = 0 \deg$, and $t = 1.0$ s, we get the maximum length of $l_{\rm ej} \sim 1.5$ km. With the observed angular distance of the ejecta ($\sim 10$ arcsec), we can calculate the distance to the event by $d \sim 20,626 \times l$. Consequently, $d$ can be constrained to be $d \lesssim 3.1 \times 10^4\ {\rm km}$.
\par
\begin{figure}
    \begin{center}
    \includegraphics[width=1.0\linewidth]{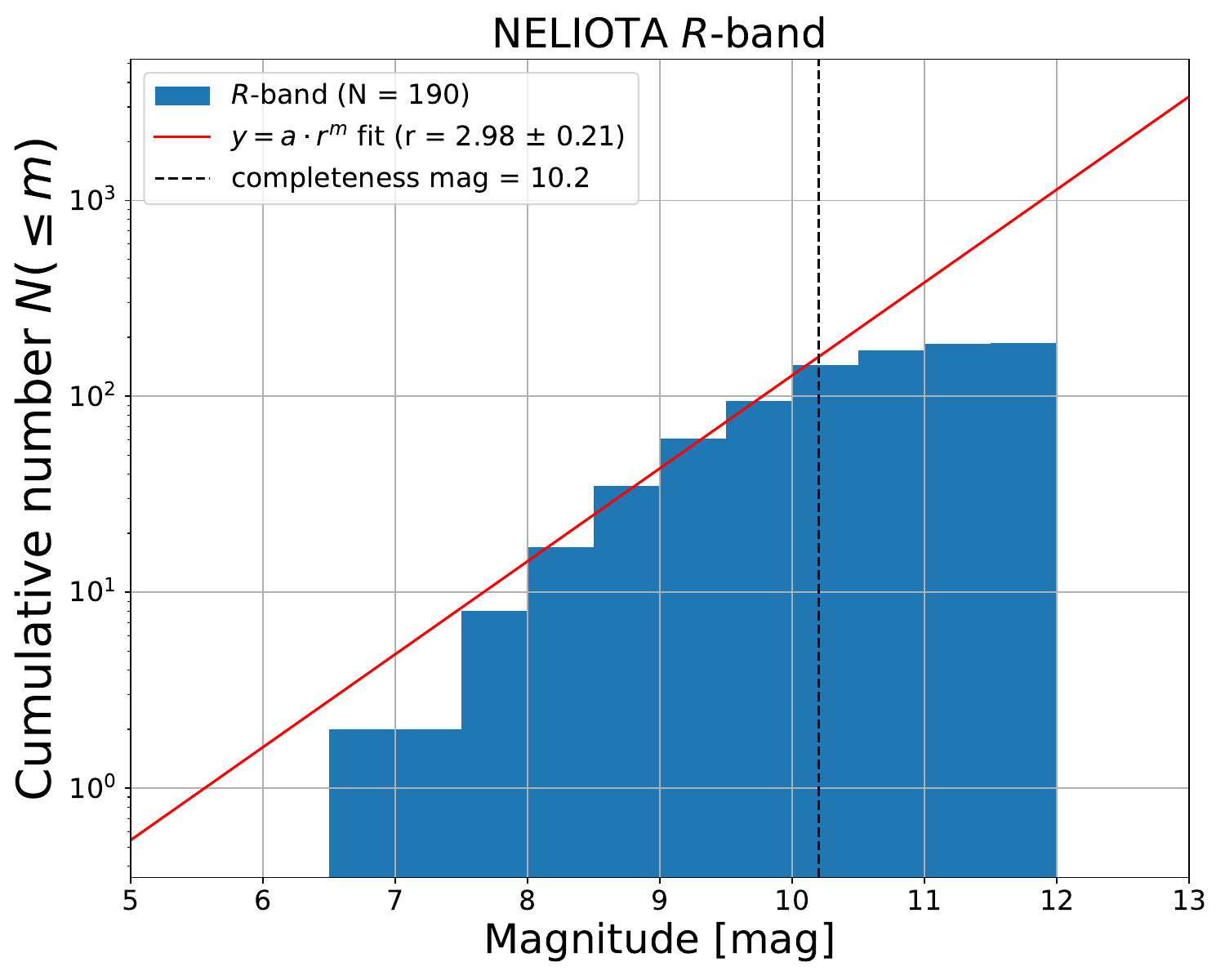}
    \end{center}
    \caption{Cumulative number distribution of the lunar impact flashes observed by NELIOTA project until June 22th, 2023. The black dashed vertical line indicates the completeness magnitude used for the fitting.
    {Alt text: Histogram of cumulative number versus magnitude with fit and reference line.}
    }
    \label{fig:neliota}
\end{figure}

The NELIOTA project \citep{Xilouris2018,Liakos2024} conducted simultaneous dual-band ($R$ and $I$) monitoring of the lunar surface to detect hundreds of lunar impact flashes. Observations covered $\sim 276$ hours until June 22, 2023, with a camera field of view (FoV) of $17'.0 \times 14'.4$. 
We fit the NELIOTA cumulative $R$-band magnitude distribution using an exponential form, where $a$ and $r$ are the fitting parameters:
\begin{equation}
    N(\leq m) = a \times r^m
\end{equation}
The best-fit parameters are $a = 0.0023 \pm 0.0016$ and $r = 2.98 \pm 0.21$ (Figure \ref{fig:neliota}). Here, $N(\leq m)$ represents the total number of impact flash events detected throughout the NELIOTA observation campaign. The total observed lunar surface area and observation time are $A_{\rm obs}=3.1\times 10^6\ {\rm km^2}$ and $T_{\rm obs} = 276\ \rm h \ (9.9 \times 10^5\ s)$, respectively.
We define the impact frequency density, $\Phi_{\rm moon}$, by normalizing the cumulative count by these parameters:
\begin{align}
    \Phi_{\rm moon}(\leq m)
    &= \frac{N(\leq m)}{A_{\rm obs} \times T_{\rm obs}} \nonumber \\
    &= \frac{0.0023\times 2.98^m}{3.1 \times 10^6 \times 9.9 \times 10^5}
       \ \left[{\rm events\ km^{-2}\ s^{-1}}\right]
\end{align}
For the detected magnitude $m\approx 16.8$:
\begin{eqnarray}
    \Phi_{\rm moon}(\leq 16.8) \sim 6.9 \times 10^{-8}\ \left[\rm events\ km^{-2}\ s^{-1}\right]
\end{eqnarray}
\par
We define the impact frequency at a distance $d$ by scaling the known lunar impact rate. Since the apparent magnitude varies with distance, we correct the magnitude threshold by the distance modulus (scaling by $5\log_{10}(d/d_{\rm moon})$). This corrected magnitude is then substituted into the lunar cumulative distribution function to obtain the expected frequency density at the target distance. For simplicity, we assume that the meteoroid flux measured on the lunar surface applies uniformly to NEA surfaces, and we ignore variations due to gravitational focusing or heliocentric orbital clustering. This assumption makes our estimate conservative and sufficient for an order-of-magnitude constraint.
%%%
\begin{align}
    \Phi(d, \leq m)
    &= \Phi_{\rm moon} \!\left(\leq m - 5\log_{10}
    \frac{d}{d_{\rm moon}} \right)
       \ \left[{\rm events\ km^{-2}\ s^{-1}}\right]
\end{align}
Table \ref{tab:scaled_frequency} shows the scaled impact frequency densities for different distances. 
%%%
\begin{table}
    \caption{Scaled impact frequency density for 16.8 mag impact flash case with various distances}
    \centering
    \label{tab:scaled_frequency}
    \begin{tabular}{@{}cc@{}}
    \hline 
    Distance [km] & $\Phi(d, \leq 16.8)$ [events\ km$^{-2}$\ s$^{-1}$] \\
    \hline 
    500 & $4.8 \times 10^{-1}$ \\
    5,000 & $2.0 \times 10^{-3}$ \\
    10,000 & $4.0 \times 10^{-4}$ \\
    31,000 & $2.7 \times 10^{-5}$ \\
    \hline
    \end{tabular}
\end{table}
%%%
%%%
We adopt the NEA population model from \citet{Harris2021} to estimate the number of NEAs; $N(D) \sim 2 \times 10^2,\ \sim 1\times 10^4$, and $\sim 1\times 10^7$ objects with diameters $D = 1.0,\ 0.1$, and $0.01~\mathrm{km}$, respectively. Assuming a uniform spatial distribution within a volume of $V \sim 1~\mathrm{AU^3}$ ($\sim 3 \times 10^{24}~\mathrm{km^3}$), each size bin yields a number density $n(D) = N(D)/V$. This assumption likely underestimates the local NEA number density near Earth, and thus makes our detection probability estimate conservative. For a given NEA size and distance, the expected number of targets crossing the FoV of Tomo-e Gozen during observations is:
\begin{equation}
    N_\mathrm{transit}(D, d) = n(D)\, V_\mathrm{survey}(d),
\end{equation}
where the survey volume swept by the FoV of Tomo-e Gozen is
\begin{equation}
    V_\mathrm{survey}(d) = w_\mathrm{FoV}^2(d)\, V_\mathrm{rel}\, T_\mathrm{obs},
\end{equation}
where $w_\mathrm{FoV}(d)$ is the linear width of the FoV at distance $d$, $V_\mathrm{rel}$ is a representative sky-plane relative velocity of NEAs (we adopt 15$~\mathrm{km\,s^{-1}}$), and $T_\mathrm{obs} = 44~\mathrm{h}$ is our total observation time. We adopt a square FoV equivalent to the total area of $17.8~\mathrm{deg^2}$. The expected number of flashes brighter than $m_\mathrm{lim}$ that occur on the surface of each transiting NEA is
\begin{equation}
    N_\mathrm{flash}(D, d) = \Phi(d, \le m_\mathrm{lim}) \,
    A_\mathrm{NEA}(D)\,
    t_\mathrm{transit}(d),
\end{equation}
where $A_\mathrm{NEA} = \pi D^2$ is the physical surface area of NEA and $t_\mathrm{transit}(d) = w_\mathrm{FoV}(d)/V_\mathrm{rel}$ is the transit timescale. The total number of detectable flashes expected during the observation is then
\begin{equation}
    N_\mathrm{expected}(D, d) = N_\mathrm{transit}(D, d)\,
    N_\mathrm{flash}(D, d).
\end{equation}
We compute $N_\mathrm{expected}(D, d)$ across the parameter space $(D,\,d)$ and present the results in Table \ref{tab:nea_expected_order}. Under the assumptions adopted, no combination of parameters satisfies $N_\mathrm{expected} \ge 1$, indicating that the probability of detecting a meteoroid impact flash on an NEA with Tomo-e Gozen during 44-hour observations is extremely small. Therefore, a NEA-meteoroid collision flash scenario of this nature cannot plausibly explain the detection of TMG20200322. 

\begin{table}[ht]
    \caption{Order-of-magnitude estimates of expected detectable flashes
    for our Tomo-e Gozen observation (44~h, $17.8~\mathrm{deg}^2$). }
    \label{tab:nea_expected_order}
    \centering
    \begin{tabular}{ccccc}
    \hline
    $D$ [km] & 500 km & 5{,}000 km & 10{,}000 km & 31{,}000 km \\
    \hline
    $0.01$ & $4\times10^{-12}$ & $2\times10^{-11}$ & $2\times10^{-11}$ & $5\times10^{-11}$ \\
    $0.1$  & $4\times10^{-13}$ & $2\times10^{-12}$ & $2\times10^{-12}$ & $5\times10^{-12}$ \\
    $1.0$  & $7\times10^{-13}$ & $3\times10^{-12}$ & $5\times10^{-12}$ & $9\times10^{-12}$ \\
    \hline
    \end{tabular}
\end{table}
%
%%%
%%%
\subsubsection{Head-on meteors}
As a potential terrestrial atmospheric origin of TMG20200322, we assume a head-on meteor. This scenario involves a meteor traveling directly towards the telescope along its line of sight, causing it to be detected as a stationary point source. Given that the coordinates and discovery time of TMG20200322 are not associated with any meteor shower, we consider meteors of sporadic origin. The typical duration of meteor luminescence ranges from $\sim 0.1 - 1.2$ s, and the fraction of sporadic
meteors with durations of $> 0.8$ s is estimated to be $\sim
8\%$ \citep{Betzler2022}. We estimate the probability of a head-on meteor with different durations of $\tau = $ 1.0 s and 0.1 s, since the exposure time of our Tomo-e Gozen observations (1.0 s) may be longer than the duration of meteors. 
In the case of a duration of 0.1 s, the peak brightness needs to be corrected for the dilution effect observed in a longer exposure time. The peak magnitude $m_{\rm P}$ is expressed as 
%%%
\begin{eqnarray}
    m_{\rm P} = -2.5\log_{10}{\left(\frac{T_{\rm exp} 10^{-0.4m_0}}{\tau}\right)},
    \label{eq:mag_dilution_correction}
\end{eqnarray}
%%%
where $\tau$ is the duration of an object and $m_0$ is the magnitude observed with an exposure time of $T_{\rm exp}$. For the observed magnitude of TMG20200322, $m_0 = 16.8$ mag (the average of two frames), the corresponding peak magnitude is $m_P = 14.3$ mag for $\tau = 0.1$ s and $T_{\rm exp} = 1.0$ s.
\par 
Here we assume that meteors occur at a constant altitude of $h = 100$ km and observations are conducted at an averaged zenith angle of $\theta = 45\deg$. 
The effective FoV of Tomo-e Gozen ($17.8\ \deg^2$) corresponds to a solid angle 
$\Omega_{\rm FoV}\approx 5.4 \times 10^{-3}$ sr. The line of sight at zenith angle $\theta$ intersects the horizontal plane at altitude $h$ along a slant distance $D = h/\cos\theta$. The area on a spherical surface at distance $D$ is $A_{\rm sphere} = \Omega_{\rm FoV} D^2$. Projecting this onto the horizontal plane, accounting for the oblique viewing angle, gives:
\begin{equation}
A = \frac{\Omega_{\rm FoV} D^2}{\cos\theta} = 
\frac{\Omega_{\rm FoV} h^2}{\cos^3\theta}.
\label{eq:projected_area}
\end{equation}
For our parameters ($h = 100$ km, $\theta = 45\deg$), this yields 
$A \approx 1.5\times 10^2\ {\rm km^2}$.
\par
We use the cumulative number flux of sporadic meteors in the $V-$band from \citet{Ohsawa2020}, fitted by an exponential form as a function of magnitude: 
\begin{eqnarray}
    F \approx 1 \times 10^{-5} \times 3.52^{M_V} \ {\rm h^{-1}\ km^{-2}},
    \label{eq:meteor_flux}
\end{eqnarray}
where $M_V$ is the absolute magnitude observed at a distance of 100 km.
We then extrapolate the relation to the fainter side to obtain the meteor flux by approximating the $V-$band and the $Gaia\ G-$band as equivalent. 
\par
For our observation geometry with a zenith angle of $45\deg$, the effective distance to the meteor layer is $D = 100/\cos(45\deg) \approx 141.4$ km. Under the approximation that the $V$-band and Gaia $G$-band are equivalent, we obtain absolute magnitudes of $M_V(\tau=1.0\ \mathrm{s}) \approx 16.0$ mag and $M_V(\tau=0.1\ \mathrm{s}) \approx 13.5$ mag for the two timescale scenarios. Extrapolating Equation~\ref{eq:meteor_flux} to these faint magnitudes and accounting for the fact that only $\sim 8\%$ of meteors persist longer than 0.8 s \citep{Betzler2022}, we calculate:
\begin{eqnarray}
    F(\tau=1.0\ {\rm s}) &\sim& 4 \times 10^{2}\ {\rm h^{-1}\ km^{-2}} \\
    F(\tau=0.1\ {\rm s}) &\sim& 2 \times 10^2\ {\rm h^{-1}\ km^{-2}}.
\end{eqnarray} 
The expected number of meteors detected in our observation $N$ can
then be estimated as the product of meteor flux, projected area at the meteor layer, and the total observation time $N = F\cdot A\cdot t_{\rm obs}$.
\begin{eqnarray}
    N(\tau = 1.0\ {\rm s}) &\sim& 3 \times 10^6\\
    N(\tau = 0.1\ {\rm s}) &\sim& 2 \times 10^6
\end{eqnarray}
For a meteor to appear as a stationary source in our sidereal tracking observations, it must move in such a way that its apparent motion cancels the telescope tracking ($15''$/s in the RA direction). The geometric probability is given by the ratio of solid angles:
\begin{eqnarray}
        p_{\mathrm{geom}} = \frac{\Omega_{\mathrm{PSF}}}{\Omega_{\mathrm{meteor}}} = \frac{\pi (\mathrm{FWHM}/2)^2}{2\pi(1-\cos\theta_{\mathrm{eff}})},
\end{eqnarray}
where $\mathrm{FWHM} = 3.0''$ is the seeing disk size, and $\theta_{\mathrm{eff}}$ is the effective zenith angle for meteor incidence. We adopt $\theta_{\mathrm{eff}} = 60\deg$ as a representative value, considering that meteors can arrive from a range of zenith angles. This yields:
\begin{equation}
    p_{\mathrm{geom}} = \frac{\pi \times (1.5'')^2 \times (\pi/648000)^2}{2\pi(1-\cos 60^\circ)} \approx 5.3 \times 10^{-11}.
\end{equation}

The expected number of head-on meteors detected in our survey $E$ is  given by $N\times p$, 
\begin{eqnarray}
    E(\tau = 1.0\ {\rm s}) &=& N(\tau = 1.0\ {\rm s}) \times p_{\mathrm{geom}} \approx 2 \times 10^{-4}\\
    E(\tau = 0.1\ {\rm s}) &=& N(\tau = 0.1\ {\rm s}) \times p_{\mathrm{geom}} \approx 1 \times 10^{-4}.
\end{eqnarray}
In both cases, the expected number of events is four orders of magnitude smaller than unity. 
In addition, such faint meteors ($M \gtrsim 14.3$ mag) have not been detected in Tomo-e Gozen 2 fps observations \citep{Ohsawa2020}. 
Extremely faint meteors are unlikely to exhibit luminescence lasting $\sim 0.1$–$1.0$~s because of their low kinetic energy. 
Therefore, it remains unlikely that TMG20200322 originates from meteors. 
%%%
%%%
\subsection{Event rate of second-timescale transients}
\begin{figure}
    \includegraphics[width=1.0\linewidth]{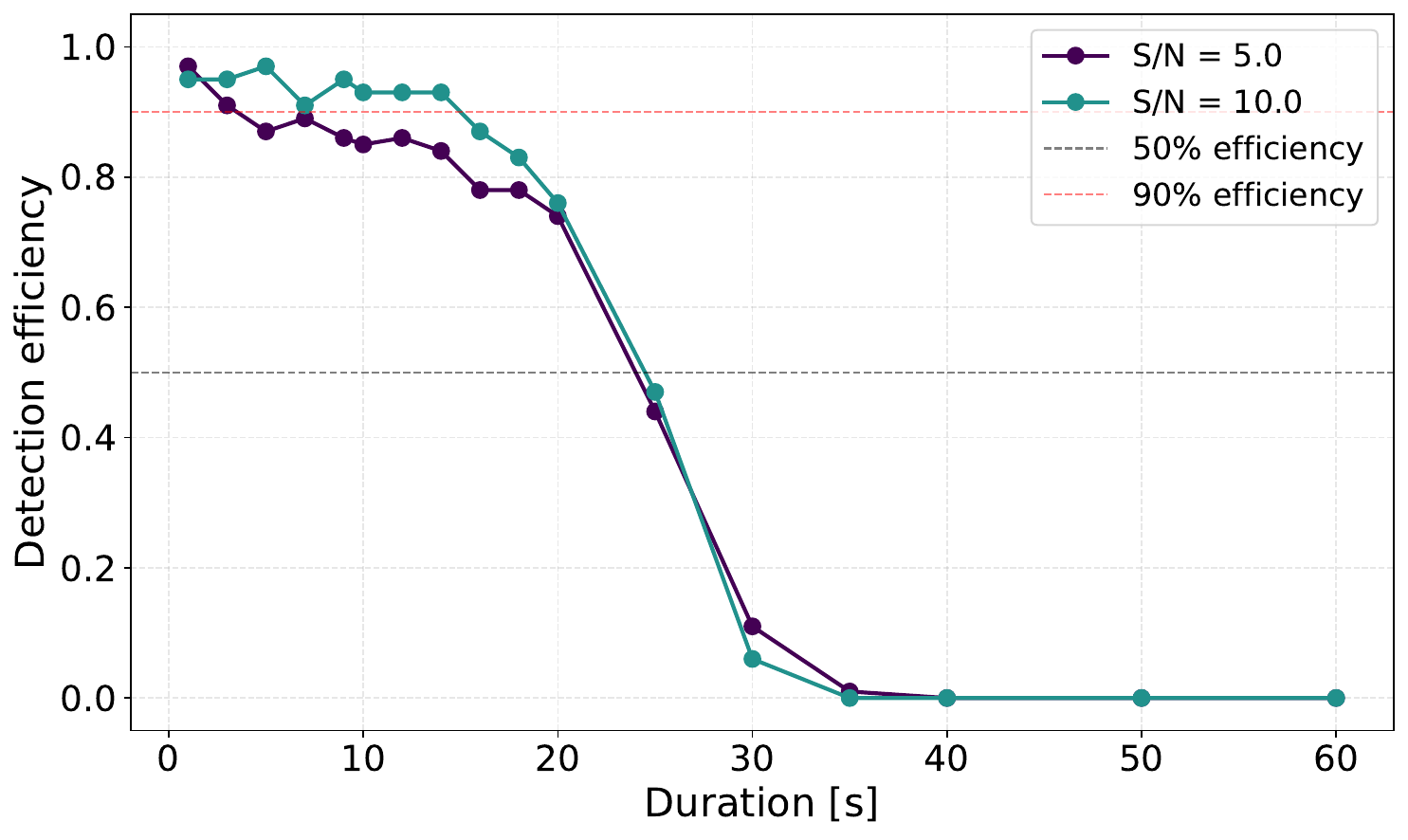}
    \caption{Detection efficiency as a function of transient duration for artificial transients with S/N = 5.0 and 10.0. The horizontal dashed lines indicate 50\% (gray) and 90\% (red) efficiency, respectively.  {Alt text: A line graph showing two curves with two reference percentage levels.}}
    \label{fig:efficieny}
\end{figure}

We discuss the event rate of second-timescale transients from the single detection of TMG20200322. We calculate the event rate of transients, $R_{\rm trans}$, in units of per square degree per day ($\deg^{-2} {\rm day^{-1}}$), as
%%%
\begin{equation}
    R_{\rm trans} = \frac{N_{\rm trans}}{\epsilon\ E_{\rm A}}
    \label{eqn:event_rate},
\end{equation}
%%%
where $N_{\rm trans}$ is the number of detected transients, and $\epsilon$ is
the detection efficiency. We perform injection-recovery tests using simulated transients (2D Gaussians) on real Tomo-e Gozen data cubes with varying durations. For each combination of S/N (5.0 and 10.0) and duration, we inject 100 synthetic transients at random spatial positions and calculate the detection efficiency as the fraction of successfully recovered events. In Figure \ref{fig:efficieny}, we find that our pipeline can reliably detect transients lasting from 1s and up to $\sim 20$ s. The measured detection efficiency is $\geq 90\%$ for transients with S/N = 10 (similar S/N to that of TMG20200322) lasting $1-15$ s. We adopt $\epsilon = 0.90$ and consider target transients with durations of 1–15 s in the following calculations.
%%%
%%%
$E_{\rm A}$ is the effective areal exposure, defined as the product of effective field of view explored and the total monitored exposure time:
%%%
\begin{equation}
    E_{\rm A} = {\rm FoV_{eff}} \times t_{\rm eff}.
\end{equation}
%%%
From the limiting magnitude distribution shown in Figure \ref{fig:limmag}, we calculate $t_{\rm eff}$ by excluding data with limiting magnitudes shallower than $m_{\rm lim} = 16$ mag, resulting in $t_{\rm eff} \sim 44.4$ hours ($\sim 1.85$ days). In our transient detection pipeline, the objects detected in the stacked images are masked, which reduces ${\rm FoV_{eff}}$ for the transient search. 
We calculate the fraction of pixels occupied by masked objects to be an average of 4.5\% of the image of $2000 \times 1128$ pixels. Furthermore, we exclude 10 pixels from the edges along both the $x$- and $y$-axes of the image for object detection, corresponding to approximately 3\% of the image area, resulting ${{\rm FoV_{eff}} = 17.78\ \deg^2}$. 
\par
The upper and lower limits for a single event ($N_{\rm trans}=1$) at the $95\%$ confidence level are $n_{\rm upper} = 4.744$ and $n_{\rm lower} = 5.13 \times 10^{-2}$ events, respectively, based on Poisson statistics \citep{Gehrels1986}. 
We take these uncertainties into account for the event rate calculation and derive 
%%%
\begin{equation}
    R_{\rm trans} = (3.4 \times 10^{-2})^{+0.13}_{-0.028} \ {\rm deg^{-2}\ day^{-1}}.
    \label{eqn:event_rate_result}
\end{equation}
%%%
The all-sky rate of FRB population is estimated to be $\sim 820$ per day above a fluence of 5 Jy ms at 600 MHz \citep{CHIMEFRB2021}. Interestingly, the all-sky rate derived from our single detection of TMG20200322 is $\sim 1,400$ per day, which is of the same order as that of the FRB population. Although the PSF elongation makes an extragalactic origin of TMG20200322 unlikely, our result suggests the possible existence of a new population of FRB-like ubiquitous optical transients.
\par
\begin{figure*}
    \begin{center}
    \includegraphics[width=1.0\linewidth]{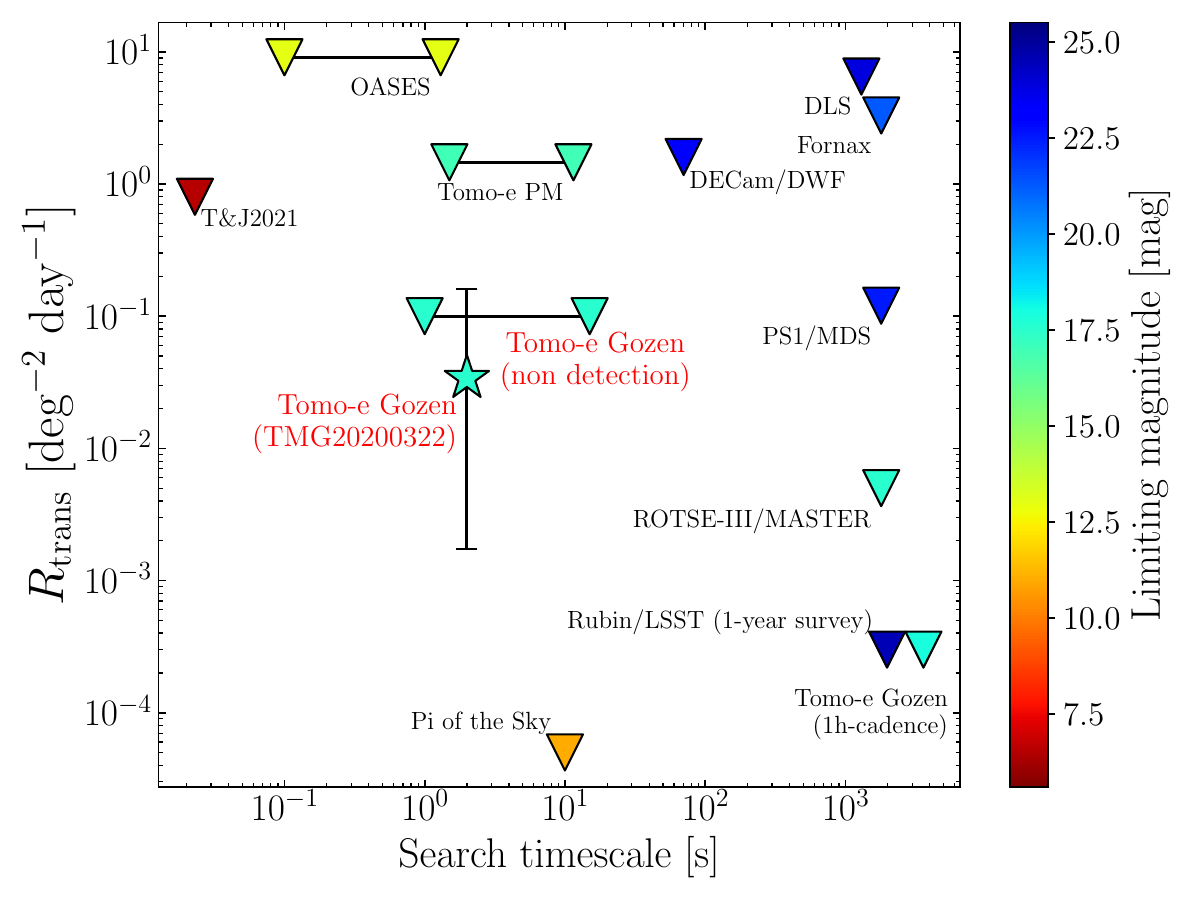}
    \end{center}
    \caption{Event rates of optical transients as a function of search timescale, compiled from short-duration (< 1 hour) transient searches in the literature. The color bar shows the limiting magnitudes of each transient searches. Our derived event rate for TMG20200322 and the upper limit for transients with $1.0~{\rm s} \leq \tau \lesssim 15~{\rm s}$ (non detection case) are indicated in red. Upper limits of the event rates ($\deg^{-2}\ {\rm day^{-1}}$) by the Tomo-e Gozen high-cadence (1h) survey \citep{Oshikiri2024}, the Deep Lens Survey (DLS; \cite{Becker2004}), the Robotic Optical Transient Search Experiment-III (ROTSE-III; \cite{Rykoff2005}), the MASTER telescope system \citep{Lipunov2007}, the ``Pi of the Sky” project \citep{Sokolowski2010}, the Fornax galaxy cluster survey \citep{Rau2008},  the Pan-STARRS1 Medium-Deep Survey (PS1/MDS; \cite{Berger2013}), the Dark Energy Camera Deeper, Wider, Faster programme (DECam/DWF; \cite{Andreoni2020}), the prototype model of the Tomo-e Gozen camera (Tomo-e PM; \cite{Richmond2020}), the Organized Autotelescopes for Serendipitous Event Survey (OASES; \cite{Arimatsu2021}) and \cite{Tingay2021} (T\&J2021) are plotted with their search timescales. The expected upper limit based on one-year operation of the Rubin/LSST main survey (g+r-band combination, 33-min cadence) \citep{Ivezic2019} is also plotted (the value is derived from \cite{Berger2013}). 
    {Alt text: Plot of event rate versus search timescale with survey markers. Inverted triangles show upper limits.}
    }
    \label{fig:event_rate}
\end{figure*}

As discussed in the previous section, the plausible astrophysical origin of TMG20200322 remains unclear. Therefore, we also estimate an upper limit on the event rate under the non-detection assumption. We follow the same procedures as described in \citet{Richmond2020} to derive an upper limit with the $95\%$ confidence level. 
Our upper limit on the event rate of transients with durations ranging from one second to less or equal to 15 seconds ($1 \rm \ s \leq \tau \lesssim 15 \rm \ s$) is estimated to be
%%%
\begin{eqnarray}
    R_{\rm trans} \lesssim 0.10\ {\rm deg^{-2}\ day^{-1}}. 
\end{eqnarray}
%%%
Figure \ref{fig:event_rate} presents our derived event rates for both the single-detection and non-detection cases, compared with values reported in previous and future transient searches targeting short timescales ($< 1$ hour). In the non-detection case, our derived upper limit constrains $\sim 14$ times tighter than that of \citet{Richmond2020}. The Pi of the Sky project obtained upper limit of an order of $10^3$ tighter than our value, but with about 6.5 mag shallower depth than Tomo-e Gozen. Survey depth is a critical factor when discussing the volumetric rate of a given class of transient sources. One of the upcoming high cadence surveys, Argus Array will provide powerful constraints on sub-second and 
short-duration transients with its unprecedented combination of cadence 
(down to 0.25 s with $\sim 15$ mag depth) and sky coverage ($\sim 8,000\ \deg^2$ ; \cite{Law2022}). Such surveys will be crucial for characterizing the population of transients like TMG20200322 and understanding their nature.
Although the distance to TMG20200322 is not constrained, surveys with shallow depth are not suitable for exploring the transient sky, particularly for sources of Galactic or extragalactic origin. 
Our possible discovery of TMG20200322 highlights that searches for second-timescale transients are highly desirable to reveal new populations of optical transients. 
%%%
%%%

\section{Summary}\label{sec:summary}
In this paper, we present a dedicated 1 fps video monitoring survey of the Earth's shadow region to probe the largely unexplored second-timescale transient sky. Using video frames with a total areal exposure of $\sim 790\ \deg^2\ {\rm hour}$, we search for objects with apparent durations of $1\leq \tau \lesssim 15$~s through our custom detection pipeline. After human visual inspection, we discover one transient candidate TMG20200322 with a duration of $\lesssim 2\ {\rm s}$.
\par
We find that TMG20200322 exhibits an elongated PSF in the second 
frame, with ellipticity ($\sim 0.50$) and semi-major axis significantly 
different from field stars. To test whether atmospheric seeing could 
produce such elongation, we perform additional observations at 
57 fps (17.4 ms exposure). Despite poorer seeing conditions in these 
data (4.0$\pm$0.3 arcsec), TMG20200322's elongation exceeds that of 
57 fps stars by 15.5$\sigma$ (semi-major axis) and 6.1$\sigma$ 
(ellipticity), demonstrating that atmospheric turbulence alone cannot 
explain the observed PSF morphology. This is consistent with simulations 
showing that $\sim$15 ms flashes remain within the seeing disk 
\citep{MegiasHomar2023}.
\par
We investigate the origins of such second-timescale transients that exhibit an elongated PSF. The expected number of events, such as impact flashes from collisions between near-Earth asteroids (NEAs) and meteoroids in the Earth's vicinity, or head-on meteors, is much less than unity and therefore cannot account for the observed occurrence rate. The derived event rate of both the single-detection and non-detection cases are compared with previous short-duration transient searches. The all-sky rate derived from the single-detection is comparable to that of FRB population, although the physical connection is need to be investigated. Even in the upper limit from the non-detection case, our constraint on the second-timescale transients is $\sim 14$ times tighter than that of initial work using the prototype model of Tomo-e Gozen. Not only wider and short exposure time but moderate limiting magnitude of $\sim 17.5$ mag, our present search can capture the transient phenomena which may be missed by shallow-depth surveys such as Pi of the Sky project \citep{Sokolowski2010}. 
\par
We have entered an era that requires optical instruments with not only wide FoV and enhanced sensitivity but also high temporal resolution. These capabilities are crucial both for unveiling the nature of known-unknown transients, such as FRBs, and for discovering new astrophysical transients in the universe.  Statistical samples of TMG20200322-like events are necessary to confirm their nature. We demonstrate that video observations using the wide-field CMOS camera targeted at the Earth's shadow region can search for transients that may have been overlooked by previous optical surveys. The upcoming operations of Rubin/LSST, with its 8.4 m diameter, will specialize in wide and deep observations \citep{Ivezic2019}. Using the wide-field, high-speed capability of Tomo-e Gozen,
we can complement Rubin/LSST by exploring the transient sky on timescales of a few seconds or less.

\begin{ack}
The authors thank the anonymous referee for his/her helpful comments that improved the manuscript. The authors also thank Dr. Tomoki Morokuma and Dr. Kazumi Kashiyama for their valuable suggestions. 
%Acknowledgement should be placed at end of main text. (NOT after the Appendix.) 
\end{ack}

\section*{Funding}
This research was supported by JSPS KAKENHI Grant Numbers JP18H01261, JP18H05223, JP21H04491, JP23K03469, JP24K17081. 

% \section*{Data availability} 
%  The data underlying this article are available ...  
% Sample Data Availability Statements 
% https://academic.oup.com/pages/open-research/research-data#Data%20Availability%20Statements

\appendix %%%%%%%%%%%%%%%%%%%%%%%%%%%%%%%%%%%%%%%%%%%%%%%%%%%%%%%%
% \section*{Case of single paragraph}
%  No section number is necessary. Add ``*'' after \verb/\section/.

%%%% 

\section*{Galactic and extragalactic origins}\label{appendix:other_origin}
The elongation of the PSF in the second frame makes TMG20200322 unlikely that this object is a Galactic or extragalactic source, but we explore possible origins, assuming that the elongation is due to extreme atmospheric turbulence. 
\par
% \subsection{Short-duration stellar flares}
As a potential source of Galactic origin, we consider a short-duration
stellar flare.  Given that TMG20200322 lacks an optical counterpart in the PS1
source catalog with a depth of $\sim 23$ mag in the $r_{\rm PS1}$-band, its
nature would be intriguing. If this is a stellar flare, the required brightening rate would exceed $\sim 6$ mag per second, equivalent to a flux increase of about 250 times. Qualitatively, the duration of stellar superflares correlates with flare energy, meaning that shorter-duration flares typically exhibit lower magnification
rates \citep{Maehara2015}. \citet{Aizawa2022} searched for short-duration stellar
flares from a one-second-cadence M-dwarf survey by Tomo-e Gozen and detected
22 flares with rise times (the time from the beginning of the stellar flare to
its peak) of ${\rm 5\ s \lesssim t_{rise} \lesssim 100\ s}$ and flare
amplitudes of $\approx 0.5 - 20$ times the quiescent luminosity. Given the relation between the rise time and amplitude of stellar flares \citep{Aizawa2022}, 
an increase in flux by a factor of more than 250 in one second is not expected unless an extreme case is considered. 
\par
\begin{figure}
    \begin{center}
    \includegraphics[width=1.0\linewidth]{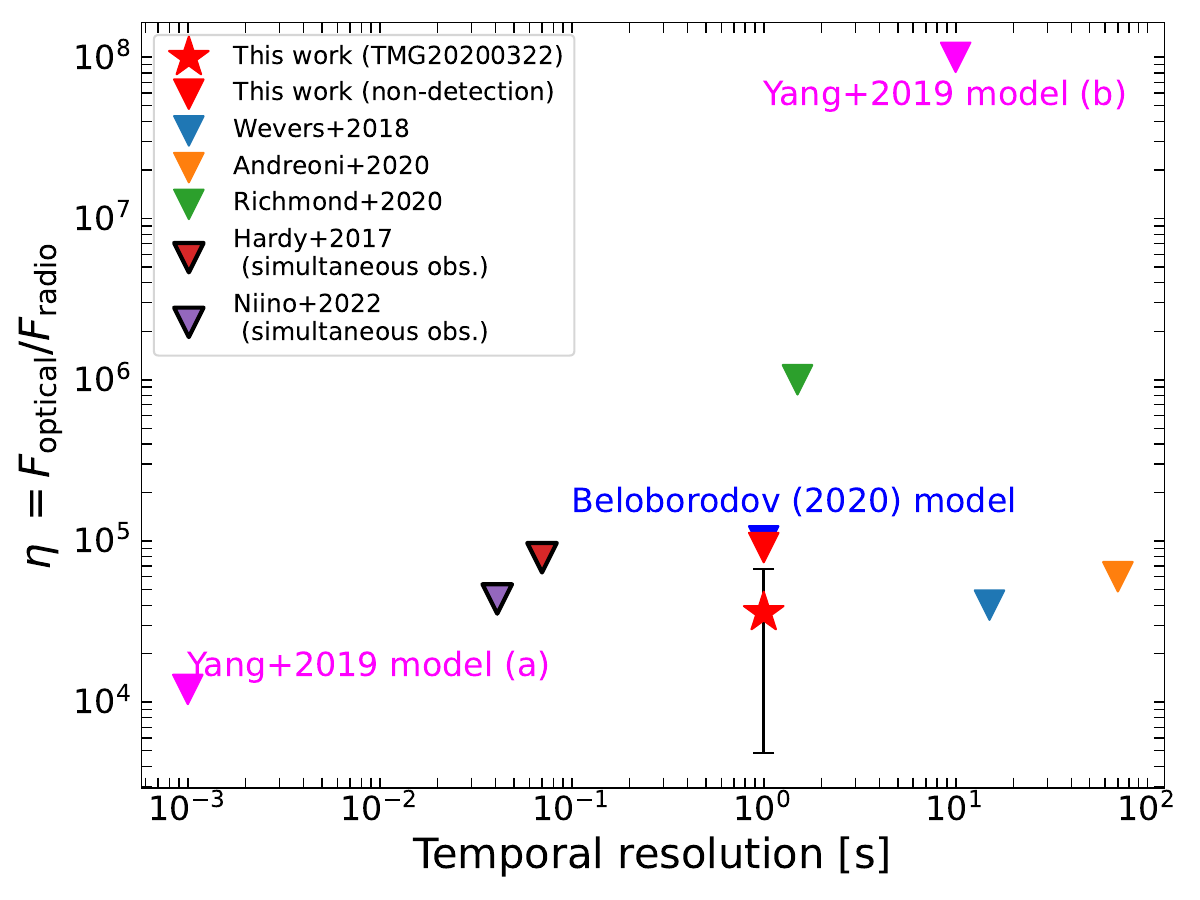}
    \end{center}
    \caption{Comparisons of the fluence ratio $\eta$ derived from several transient searches which are summarized by \citet{Chen2020} (\cite{Wevers2018,Andreoni2020,Richmond2020}) as a function of the temporal resolution of each study. We include results from the simultaneous observations of FRBs with optical and radio bands, and model predictions of FRB optical emissions. The error bar of the point (detection case) indicates the $3\sigma$ of the power law index of the radio fulence CDF model \citep{James2019}.
    {Alt text: Fluence ratio vs temporal resolution with study markers, upper-limit triangles, a red star for this work, and model labels.}
    }
    \label{fig:frb_eta}
\end{figure}

% \subsection{Optical counterpart to FRBs}
As a transient of extragalactic origin, here we assume that TMG20200322 is an 
optical counterpart to FRBs. Although simultaneous observations using radio and optical bands \citep{Hardy2017,Niino2022,Hiramatsu2023}, and follow-up observations by multiple wavelengths have been conducted within hours of a burst \citep{Petroff2015, Tominaga2018}, there has been no detection
of optical emission from any FRB. Interestingly, a model involving a magnetar
giant flare producing an FRB predicts an optical flash with duration of $\sim
0.1$ to $\sim 1\ {\rm~s}$ \citep{Beloborodov2020}. The luminosity is expected to
be $L_{\rm iso,\ peak} \sim 10^{44 - 45}\ {\rm erg/s}$, occurring synchronously
with an FRB \citep{Beloborodov2020}. We follow the same technique by \citet{Chen2020} to calculate the fluence ratio of the optical and radio bands of FRBs using a cumulative fluence distribution derived from independent radio observations. 
\begin{equation} \eta(\nu_{c}) \equiv \frac{F_{\rm optical}}{F_{\rm radio}} =
\frac{\nu_{1,\ \rm optical} \cdot F_{\nu,\ \rm optical}}{\nu_{1,\ \rm radio}
\cdot F_{\nu,\ \rm radio}},\label{eq:integ_fluence_ratio} \end{equation}
where the central frequency $\nu_{c}$ for Tomo-e Gozen is $\nu_{c} = 6.0 \times
10^{14}$ Hz ($\lambda_{\rm c} = 500\ {\rm nm}$). $F_{\rm optical}$ and $F_{\rm
radio}$ are the band-integrated fluences, $F_{\nu,\ \rm optical}$ and
$F_{\nu_{1,\ \rm radio}}$ are specific fluences, and $\nu_{1,\ \rm optical}$
and $\nu_{1,\ \rm radio}$ are the lower frequency bounds for optical and radio
bands, respectively. This method allows us to test whether our assumption of TMG20200322 to be an optical counterpart to FRBs is valid or inconsistent. Although the origin of TMG20200322 remains unidentified, the derived $\eta (\nu_{c})$ is consistent with the theoretical model for FRBs which involves a magnetar giant flare \citep{Beloborodov2020} (Figure \ref{fig:frb_eta}). 
%%%
%%%

% Any journal's BST file (e.g., apj.bst) can be used as PASJ's BST is unavailable.    
% \bibliographystyle{****}
% \bibliography{****}
\bibliographystyle{aasjournal}
\bibliography{tomoe_flash_pasj}

\end{document}